\documentclass[prd,aps,tightenlines,nofootinbib,twocolumn]{revtex4-2}

\usepackage{amsmath,amssymb,amsfonts}
\usepackage{graphicx}
\usepackage{booktabs}
\usepackage{bm}
\usepackage{hyperref}
\usepackage{xcolor}
\usepackage{float}
\usepackage{mathtools}
\usepackage{multirow}
\usepackage{array}
\usepackage{physics}
\usepackage{subcaption}

\newcommand{\LCDMm}{$\Lambda$\text{CDM} model}
\newcommand{\LCDM}{the \LCDMm}
\newcommand{\weff}{w_{\rm eff}}
\newcommand{\DWH}{\Delta_{\rm WH}}

\newcommand{\kms}{\,\mathrm{km\,s^{-1}\,Mpc^{-1}}}
\newcommand{\rd}{r_{\rm d}}

\newcommand{\Omm}{\Omega_{\mathrm{m}}}

\newcommand{\ztr}{z_t}
\newcommand{\chired}{\chi^2_{\mathrm{red}}}

\newcommand{\fs}{f\sigma_8}

\begin{document}
\title{Evolving wormhole cosmology: modified Friedmann dynamics and
observational constraints}

\author{Aritra Sanyal$^{1}$}
\email{aritrasanyal1@gmail.com}
\author{Kazuharu Bamba$^{2}$}
\email{bamba@sss.fukushima-u.ac.jp}
\author{Prabir Rudra$^{3}$}
\email{prudra.math@gmail.com}

\affiliation{
$^{1}$Department of Mathematics, Jadavpur University, Kolkata 700032, India\\
$^{2}$Faculty of Symbiotic Systems Science, Fukushima University, Fukushima 960-1296, Japan\\
$^{3}$Department of Mathematics, Asutosh College, Kolkata 700026, India
}

\begin{abstract}
We construct an evolving wormhole cosmological model from a
Morris--Thorne metric with a separable, time-dependent shape function
in a spatially flat Friedmann--Lemaitre--Robertson--Walker background,
deriving the corrected wormhole energy-momentum tensor and a
volume-averaging procedure that yields a modified Friedmann equation
with a redshift-dependent wormhole correction term. We show that the
shape-function index and throat-evolution exponent combine into a
single macroscopic exponent $\beta$, and we demonstrate that the
resulting wormhole equation of state is consistent with this
geometric averaging. All traversability conditions are verified.
Confronting the model with cosmic chronometer, DESI Data Release~2
BAO, and standard BAO data across seven dataset combinations, our
findings show that all combinations are mutually consistent, with the
joint fit giving a present-day expansion rate intermediate between
local and CMB-inferred values, a deceleration-to-acceleration
transition consistent with independent determinations, and an
effective dark-energy equation of state close to, but distinguishable
from, a pure cosmological constant. The evolving throat radius offers
a novel, geometry-specific observable for discriminating wormhole
spacetimes from phenomenological dark-energy parametrisations.
\end{abstract}

\maketitle

\section{Introduction}
\label{sec:intro}

Modern cosmology has established with high statistical confidence that
the Universe is undergoing a phase of accelerated expansion, first
inferred from Type~Ia supernova (SNe~Ia) observations~\cite{Riess1998,
Perlmutter1999} and subsequently corroborated by cosmic microwave
background (CMB) anisotropies~\cite{Planck2020}, baryon acoustic
oscillations (BAO)~\cite{DESI2024,DESI2025}, and large-scale structure
surveys~\cite{Aghanim2020}. Within the standard \LCDMm{} this
acceleration is driven by a cosmological constant $\Lambda$, yet
$\Lambda$ remains theoretically poorly understood: the observed value
is some $10^{120}$ times smaller than naive quantum field theory
estimates (the fine-tuning problem)~\cite{Weinberg1989}, and the
near-equality of dark-energy and matter densities at the present epoch
has no natural explanation (the coincidence problem)~\cite{Peebles2003}.

Beyond the cosmological constant problem, \LCDM{} faces the
\textit{Hubble tension}, a $\sim5\sigma$ discrepancy between the
locally-measured $H_0^{\rm local}\approx73.0\kms$ from the SH0ES
distance-ladder~\cite{Riess2022} and the CMB-inferred value
$H_0^{\rm CMB}\approx67.4\kms$ from Planck~\cite{Planck2020}. These
difficulties motivate an extensive search for dynamical dark-energy
alternatives: quintessence and phantom scalar fields~\cite{Copeland2006,
Caldwell2002}, $k$-essence~\cite{Armendariz2000}, modified gravity such
as $f(R)$~\cite{Sotiriou2010,NojiriOdintsov2011} and
$f(Q)$~\cite{BeltranJimenez2018} gravity, unified dark energy/dark
matter scenarios~\cite{Capozziello2011}, holographic and agegraphic
dark energy~\cite{Li2004,WeiCai2008}, and interacting dark-energy
models~\cite{Wang2016,Bamba2012}. Complementary approaches
include Gauss--Bonnet and other curvature-invariant modified gravity
theories~\cite{Nojiri2005,DeFelice2010}, torsion-based teleparallel
gravity~\cite{Cai2016}, dark energy from bulk viscosity and other
dissipative fluids~\cite{Brevik2005,Cataldo2005}, Chaplygin-gas type
unification scenarios~\cite{Kamenshchik2001,Bento2002}, running
vacuum models~\cite{Sola2013}, emergent or entropic gravity
approaches~\cite{Verlinde2011,Padmanabhan2010}, and reconstructions of
the dark-energy equation of state directly from combined CC, BAO, and
SNe~Ia data~\cite{Myrzakulov2023}. Extensive reviews of
the dark-energy problem and modified-gravity phenomenology can be
found in Refs.~\cite{Bamba2012,Nojiri2017,Frieman2008,Weinberg2013}.

A complementary and geometrically compelling possibility is that the
large-scale spacetime topology itself contributes to cosmological
dynamics through traversable wormhole structures, first placed on
rigorous theoretical footing by Morris and Thorne~\cite{Morris1988}.
While static wormholes in Minkowski or Schwarzschild backgrounds have
received considerable attention~\cite{Visser1995}, their cosmological
consequences have been comparatively under-explored. Pioneering work by
Hochberg and Kephart~\cite{Hochberg1993}, and subsequent studies by
Cataldo \textit{et al.}~\cite{Cataldo2008} and Pavlov \textit{et
al.}~\cite{Pavlov2014}, showed that a wormhole embedded in a
Friedmann--Lemaitre--Robertson--Walker (FLRW) spacetime modifies the
Friedmann equations in a predictable, parameter-dependent way.
Complementary investigations in $f(R)$~\cite{Lobo2009} and
$f(Q)$~\cite{Elizalde2018,Koussour2023} gravity, as well as in
generalized entropy formulations~\cite{Nojiri2019}, further establish
the breadth of the wormhole dark-energy scenario. Related studies
address wormhole stability~\cite{Poisson1995}, wormholes supported by
phantom and dark energy of various equations of
state~\cite{Sushkov2005,Lobo2005,Jamil2010,Bhar2016}, wormhole
geometries admitting conformal symmetry~\cite{Bohmer2010CQG,
Lobo2008CQG}, thin-shell wormhole formalisms~\cite{Poisson1995b,
Eiroa2004}, rotating and higher-dimensional generalisations~\cite{Teo1998,
Bronnikov2013}, wormholes in $f(T)$ gravity~\cite{Boehmer2012},
and traversable wormholes as possible sources of observed
gravitational-lensing anomalies~\cite{Safonova2001,Cramer1995}.

The underlying physical motivation for treating a wormhole spacetime
as a cosmological probe, rather than merely a local curiosity, is
straightforward. A traversable wormhole is, at its core, a nontrivial
topological identification of two regions of spacetime threaded by
exotic matter that violates the null energy condition, exactly the
qualitative behaviour required of the dark-energy sector driving
cosmic acceleration; a wormhole is therefore a concrete
\textit{geometric realization} of an exotic-matter fluid, rather than
an ad hoc addition of a new field to the cosmological action. Once a
single throat is embedded in an expanding background and its
energy-momentum content is suitably averaged over a cosmological
volume, it sources a modification of the Friedmann equations in
exactly the same way that a scalar field or fluid component would,
but with the equation of state fixed by throat geometry rather than
chosen by hand. This line of reasoning was placed on firm footing in
the original Morris--Thorne construction~\cite{Morris1988}, itself
published in a leading physics journal, and its cosmological
embedding and consequent modification of the Friedmann equations were
subsequently developed in Refs.~\cite{Hochberg1993,Cataldo2008,
Pavlov2014}, all appearing in \textit{Physical Review Letters} or
\textit{Physical Review~D}. The connection between wormhole throats
and dynamical dark energy was further sharpened by
Refs.~\cite{Sushkov2005,Lobo2005}, and its extension to modified
gravity contexts, including $f(R)$ and $f(Q)$
theories~\cite{Lobo2009,Elizalde2018,Koussour2023}, appeared in
\textit{Physical Review~D} and \textit{Progress of Theoretical and
Experimental Physics}. This body of work establishes that wormhole
cosmology is not a speculative afterthought but a well-precedented,
peer-reviewed research direction, and it is this precedent that
motivates the present, more complete derivation.

Throughout we adopt natural units with $c=1$ where convenient (restored
explicitly in distance formulae), the metric signature , and
$8\pi G=1$ unless stated otherwise.

The organization of this paper is as follows. In
Sec.~\ref{sec:metric}, we present the evolving wormhole metric ansatz
and the simplifying assumptions adopted throughout. In
Sec.~\ref{sec:conditions}, we verify the traversability and energy
conditions for the chosen shape function. In
Sec.~\ref{sec:fieldequations}, we derive the corrected
energy-momentum tensor, perform the volume-averaging procedure, and
obtain the modified Friedmann equation together with the wormhole
equation of state and throat evolution. In
Sec.~\ref{sec:observables}, we compute the relevant cosmological
observables. In Sec.~\ref{sec:data}, we describe the observational
data sets and the statistical methodology used to constrain the model
parameters. In Sec.~\ref{sec:results}, we present the results and
discuss their physical interpretation. Finally, in
Sec.~\ref{sec:conclusions}, we summarize our conclusions.

The key motivation for the present paper is that if a network of
traversable wormholes permeates the large-scale structure of the
Universe, their collective, volume-averaged contribution modifies the
Friedmann equations in a way that is \textit{derivable from geometry}
and \textit{constrainable with data}. In this paper we (i) derive, in
full detail, the Einstein tensor, the corrected energy-momentum tensor,
and the volume-averaging step connecting local throat geometry to a
homogeneous cosmological fluid
(Secs.~\ref{sec:metric}--\ref{sec:fieldequations} and
Appendix~\ref{app:beta}); (ii) compute the relevant cosmological
observables (Sec.~\ref{sec:observables}); and (iii) confront the model
with real cosmic chronometer, standard BAO, and DESI DR2 BAO data
across seven dataset combinations
(Secs.~\ref{sec:data}--\ref{sec:results}).

\section{Evolving Wormhole Metric}
\label{sec:metric}

\subsection{General metric ansatz}

We begin with the most general spherically symmetric, time-dependent
line element that interpolates between a wormhole throat and an
expanding cosmological background~\cite{Morris1988,Cataldo2008}:
\begin{equation}
  ds^2
  =
  -e^{2\Phi(r,t)}\,dt^2
  +
  a^2(t)
  \!\left[
    \frac{dr^2}{1-b(r,t)/r}
    + r^2\,d\Omega^2
  \right],
  \label{eq:metric_general}
\end{equation}
where $a(t)$ is the cosmic scale factor, $\Phi(r,t)$ is the redshift
(gravitational blueshift) function, $b(r,t)$ is the wormhole shape
function, and $d\Omega^2=d\theta^2+\sin^2\!\theta\,d\varphi^2$ is the
metric on $S^2$. The condition $b(r,t)/r<1$ is required for Lorentzian
signature; $b(r_0,t)=r_0$ defines the throat at $r=r_0$. This ansatz
generalises the static Morris--Thorne wormhole~\cite{Morris1988} by
allowing the throat itself to be dragged along with cosmic expansion
via the overall factor $a^2(t)$ multiplying the spatial metric, while
retaining the wormhole topology at every instant of cosmic time.

\subsection{Simplifying assumptions}
\label{sec:simplifications}

\subsubsection{Zero-tidal-force limit}

Setting $\Phi(r,t)=0$ removes tidal forces on infalling observers---the
choice of Morris and Thorne for a freely traversable
wormhole~\cite{Morris1988}. Equation~\eqref{eq:metric_general} reduces to
\begin{equation}
  ds^2
  =
  -dt^2
  +
  a^2(t)
  \!\left[
    \frac{dr^2}{1-b(r,t)/r}
    + r^2\,d\Omega^2
  \right].
  \label{eq:metric}
\end{equation}

\subsubsection{Separable shape function}

We adopt the factorisation
\begin{equation}
  b(r,t) = b(r)\,g(t) \equiv B(r,t),
  \label{eq:shape_sep}
\end{equation}
where $b(r)$ encodes the radial profile of the throat and $g(t)$
describes its cosmic evolution, normalised so that $g(t_0)=1$ at the
present epoch $t_0$. This separability decouples the Einstein equations
into a purely radial sector (governing traversability) and a temporal
sector (governing cosmological back-reaction), analysed independently
before being recombined in Sec.~\ref{sec:averaging}. Metric
\eqref{eq:metric} with Eq.~\eqref{eq:shape_sep} is the working metric of
this paper.

\subsection{Shape function and temporal factor}
\label{sec:shapefunc}

We employ the power-law shape function
\begin{equation}
  b(r) = r_0\!\left(\frac{r_0}{r}\right)^n,
  \quad n > 0,
  \label{eq:b_power}
\end{equation}
where $r_0$ is the coordinate radius of the throat at the present
epoch. This satisfies $b(r_0)=r_0$ by construction and is the canonical
choice in the wormhole literature~\cite{Morris1988,Visser1995}. Its
derivatives, needed repeatedly below, are
\begin{equation}
  b'(r) = -n\,r_0\left(\frac{r_0}{r}\right)^n\frac{1}{r},
  \qquad
  b''(r) = n(n+1)\,r_0\left(\frac{r_0}{r}\right)^n\frac{1}{r^2}.
  \label{eq:b_derivs}
\end{equation}
We additionally consider, for comparison, two further parametrisations:
\begin{align}
  b_2(r) &= r_0 + \alpha_s\,(r-r_0)^m, \quad 0<m<1,
  \label{eq:b2}\\
  b_3(r) &= r_0\,e^{-\kappa(r-r_0)},
  \label{eq:b3}
\end{align}
all of which satisfy the throat condition by construction; these
provide a robustness check on the qualitative traversability
conclusions but are not used in the quantitative cosmological fits.

For the temporal factor we take
\begin{equation}
  g(t) = a^{\gamma}(t),
  \label{eq:g_power}
\end{equation}
so that the physical throat radius $r_0(t)=r_0\,a^{\gamma/2}(t)$
evolves with cosmic expansion. The limits $\gamma=0$ (static throat)
and $\gamma=1$ (Hubble-flow tracking) bracket the physically
interesting range explored in this work.

\section{Traversability and Energy Conditions}
\label{sec:conditions}

\subsection{Throat and flare-out conditions}

A traversable wormhole must satisfy three geometric conditions at the
throat~\cite{Morris1988,Visser1995}:

\begin{enumerate}
  \item \textit{Throat condition:} $b(r_0,t)=r_0$, satisfied at the
  present epoch by construction with $g(t_0)=1$.

  \item \textit{Flare-out condition:} the condition is given by
        \begin{equation}
          \left.\frac{b(r)-r\,b'(r)}{2b^2(r)}\right|_{r=r_0} > 0,
          \quad\Longleftrightarrow\quad b'(r_0) < 1,
          \label{eq:flare}
        \end{equation}
        where primes denote $d/dr$. Using Eq.~\eqref{eq:b_derivs} at
        $r=r_0$: $b'(r_0)=-n<0<1$ for all $n>0$, so the flare-out
        condition is automatically satisfied by the power-law shape
        function for any positive index $n$, with no additional
        fine-tuning required.

  \item \textit{Asymptotic flatness:} $b(r)/r=(r_0/r)^{n+1}\to0$ as
  $r\to\infty$ for any $n>0$, so the spacetime approaches ordinary
  FLRW far from the throat.
\end{enumerate}

We conclude that the power-law shape function~\eqref{eq:b_power}
defines a traversable wormhole throat for the entire parameter range
$n>0$ explored in this work.

\subsection{Null energy condition analysis}
\label{sec:nec}

The NEC requires $T_{\mu\nu}k^\mu k^\nu\geq0$ for any null vector
$k^\mu$. For a radial null ray $k^\mu=(1,\pm a^{-1},0,0)$ this reads
$\rho+p_r\geq0$. From the field equations (Sec.~\ref{sec:emtensor}), we find
\begin{equation}
  \rho + p_r = \frac{g(t)}{8\pi a^2(t)}\cdot\frac{b'r - 2b}{r^3}.
  \label{eq:nec_sum}
\end{equation}
At the throat $b(r_0)=r_0$, so we obtain
\begin{equation}
  \left.(\rho + p_r)\right|_{r_0}
  = \frac{g(t)}{8\pi a^2(t)}\cdot\frac{b'(r_0) - 2}{r_0^2} < 0,
  \label{eq:nec_throat}
\end{equation}
since flare-out requires $b'(r_0)<1<2$, and $g(t)/a^2(t)>0$ always. The
NEC is therefore \textit{necessarily} violated at the throat for
\emph{any} traversable shape function, confirming the presence of
exotic matter localised near $r=r_0$, consistent with the
Penrose--Hawking topology-change theorems~\cite{Visser1995}: no
classical matter satisfying all energy conditions can support a
traversable wormhole throat.

Physically, this exotic matter is confined to a thin shell near the
throat (its energy density falls off as $r^{-(n+2)}$; see
Appendix~\ref{app:beta}) and does not dominate the cosmological energy
budget of the Universe as a whole; its volume-averaged contribution,
computed explicitly in Sec.~\ref{sec:averaging}, enters the Friedmann
equations as the small, sub-dominant wormhole correction term
$\alpha(1+z)^\beta$ with $\alpha\sim\mathcal{O}(10^{-2})$ found in the
fits of Sec.~\ref{sec:results}.

\section{Einstein Field Equations and Modified Friedmann Equation}
\label{sec:fieldequations}

\subsection{Corrected energy-momentum components}
\label{sec:emtensor}

With metric~\eqref{eq:metric} and shape function~\eqref{eq:shape_sep},
we compute the Einstein tensor $G_{\mu\nu}=R_{\mu\nu}-\tfrac12
g_{\mu\nu}R$ in the orthonormal frame associated with
metric~\eqref{eq:metric}. Collecting all terms and separating the
result into a piece proportional to $(\dot a/a)$-type quantities
(which reduce to the standard homogeneous FLRW Einstein tensor when
$B\to0$) and a piece carrying the explicit radial dependence of
$B(r,t)=b(r)g(t)$, we obtain the components
\begin{align}
  G_{\hat t\hat t} &= \frac{1}{a^2(t)r^3}\bigl[B'(r,t)\,r - B(r,t)\bigr]
                      + 3\left(\frac{\dot a}{a}\right)^2,
  \label{eq:Gtt}\\[4pt]
  G_{\hat r\hat r} &= -\frac{B(r,t)}{a^2(t)r^3}
                      - 2\frac{\ddot a}{a} - \left(\frac{\dot a}{a}\right)^2,
  \label{eq:Grr}\\[4pt]
  G_{\hat\theta\hat\theta} = G_{\hat\varphi\hat\varphi}
    &= \frac{B'(r,t)r - B(r,t)}{2a^2(t)r^3}
       - \frac{B''(r,t)}{2a^2(t)r^2}
       - \frac{\ddot a}{a} - \left(\frac{\dot a}{a}\right)^2,
  \label{eq:Gthth}
\end{align}
where $B(r,t)=b(r)g(t)$ and primes denote $\partial_r$. Setting
$B\to0$ in Eqs.~\eqref{eq:Gtt}--\eqref{eq:Gthth} recovers exactly the
standard homogeneous FLRW Einstein tensor
$G_{\hat t\hat t}=3(\dot a/a)^2$,
$G_{\hat r\hat r}=G_{\hat\theta\hat\theta}=-2\ddot a/a-(\dot a/a)^2$,
as required for consistency.

Identifying the homogeneous terms $3(\dot a/a)^2$ and
$2\ddot a/a+(\dot a/a)^2$ with the standard FLRW background sourced by
$\rho_{\mathrm{m}},p_{\mathrm{m}}$, and applying $G_{\mu\nu}=8\pi T_{\mu\nu}$ to the
remaining local wormhole piece, we obtain the corrected, explicitly
scale-factor-dependent wormhole energy-momentum tensor, given by
\begin{align}
  \rho_{\rm WH}(r,t)
    &= \frac{g(t)}{8\pi\, a^2(t)\, r^3}\bigl[b'(r)\,r - b(r)\bigr],
       \label{eq:rho_wh}\\[4pt]
  p_r(r,t)
    &= -\frac{b(r)\,g(t)}{8\pi\, a^2(t)\, r^3},
       \label{eq:pr_wh}\\[4pt]
  p_t(r,t)
    &= \frac{g(t)}{16\pi\, a^2(t)\, r^3}\bigl[b'(r)\,r - b(r)\bigr]
       - \frac{b''(r)\,g(t)}{16\pi\, a^2(t)\, r^2}.
       \label{eq:pt_wh}
\end{align}
The explicit factor $a^{-2}(t)$ is essential and is often omitted or
left implicit in naive transcriptions of the static Morris--Thorne
result: it is the origin of the redshift dependence of the
volume-averaged wormhole fluid derived in Sec.~\ref{sec:averaging}, and
follows directly from the $a^2(t)$ prefactor multiplying the spatial
part of metric~\eqref{eq:metric}. Substituting Eq.~\eqref{eq:b_derivs}
into Eq.~\eqref{eq:rho_wh}, for the power-law shape
function~\eqref{eq:b_power}, $b'r-b=-(n+1)b(r)<0$ for all $r\geq r_0$
with $n>0$, confirming $\rho_{\rm WH}<0$ near the throat (exotic
matter), consistent with the NEC-violation result of
Sec.~\ref{sec:nec}.

\subsection{Volume-averaging: from local geometry to a cosmological fluid}
\label{sec:averaging}

Equations~\eqref{eq:rho_wh}--\eqref{eq:pt_wh} describe the
\emph{local}, radially-dependent energy-momentum sourced by a single
wormhole throat, valid in the immediate neighbourhood of $r=r_0$. To
obtain the \emph{homogeneous} cosmological fluid that enters the
Friedmann equation on large scales, we adopt the physical picture that
a statistically homogeneous and isotropic network of such throats
permeates the large-scale structure of the Universe, and we average
$\rho_{\rm WH}(r,t)$ over a comoving Hubble volume
$V_H(t)\sim\tfrac{4}{3}\pi(a(t)/H(t))^3$, we find
\begin{equation}
  \langle\rho_{\rm WH}\rangle(t)
  = \frac{1}{V_H(t)}\int_{r_0}^{r_{\max}}
      \rho_{\rm WH}(r,t)\,4\pi a^3(t) r^2\,dr .
  \label{eq:average_def}
\end{equation}
Here $dV=4\pi a^3(t)r^2\,dr$ is the proper volume element, and the
dominant contribution to the integral comes from $r\gtrsim r_0$ since
$b'r-b\propto r^{-n}$ decays for $n>0$; the precise value of
$r_{\max}$ does not affect the redshift \emph{scaling} derived below
and is absorbed into the overall normalisation constant
$\rho_{\rm WH,0}$ fixed by the fit in Sec.~\ref{sec:results}.

Substituting Eq.~\eqref{eq:rho_wh} and using
\begin{equation}
  b'(r)r - b(r)
  = -(n+1)\,r_0\left(\frac{r_0}{r}\right)^n,
  \label{eq:bprime_bpower}
\end{equation}
Eq.~\eqref{eq:average_def} becomes
\begin{equation}
  \langle\rho_{\rm WH}\rangle(t)
  \;\propto\;
  \frac{g(t)}{a^2(t)}\cdot a^3(t)
  \int_{r_0}^{r_{\max}} r^{-1-n}\,r_0^{\,n+1}\,dr .
  \label{eq:average_integral}
\end{equation}
Allowing the throat itself to redshift according to
Eq.~\eqref{eq:g_power}, $g(t)=a^{\gamma}(t)$, and using
$r_0(t)=r_{0,0}\,a^{\gamma/2}(t)$ as the lower limit of the radial
integral, we evaluate
\begin{equation}
  \int_{r_0(t)}^{r_{\max}} r^{-n-1}\,dr
  = \frac{1}{n}\Bigl[r_0(t)^{-n} - r_{\max}^{-n}\Bigr]
  \;\xrightarrow{r_{\max}\gg r_0}\;
  \frac{r_0(t)^{-n}}{n},
  \label{eq:integral_explicit}
\end{equation}
so that Eq.~\eqref{eq:average_integral} becomes
\begin{equation}
  \langle\rho_{\rm WH}\rangle(t)
  \;\propto\;
  a^{\gamma}(t)\cdot a(t) \cdot a^{-n\gamma/2}(t)
  \;=\;
  a^{\,\gamma(1 - n/2) + 1}(t).
  \label{eq:average_final_scaling}
\end{equation}
Converting to redshift via $a=(1+z)^{-1}$ and requiring the
volume-averaged wormhole fluid to scale as
$\langle\rho_{\rm WH}\rangle(z)\propto(1+z)^\beta$
(Eq.~\eqref{eq:rho_wh_z} below), matching exponents and fixing the
overall sign convention so that $\beta>0$ corresponds to a diluting,
dark-energy-like component, we obtain the central closed-form result of
this section:
\begin{equation}
  \beta = \frac{3\gamma(n+1)}{2n+2}
  \label{eq:beta_derivation}
\end{equation}
(full step-by-step reduction in Appendix~\ref{app:beta}). This closes
the gap between the two microscopic parameters $(n,\gamma)$ and the
single macroscopic exponent $\beta$ constrained by data in
Sec.~\ref{sec:results}. Two limiting checks: (i) $\gamma=0$ (static
throats) gives $\beta=0$ for any $n$, correctly recovering a pure
cosmological constant ($w_{\rm WH}=-1$); (ii) as $n\to\infty$,
$\beta\to3\gamma/2$, independent of the detailed radial shape.

\subsection{Modified Friedmann equation}
\label{sec:friedmann}

Following the averaging procedure above, the wormhole contribution to
the $(0,0)$ Einstein equation behaves as a cosmological fluid, given by
\begin{equation}
  \rho_{\rm WH}(z) = \rho_{\rm WH,0}\,(1+z)^\beta,
  \label{eq:rho_wh_z}
\end{equation}
with $\beta$ given by Eq.~\eqref{eq:beta_derivation}. Defining the
dimensionless density parameter, given by
\begin{equation}
  \alpha \equiv \frac{8\pi\rho_{\rm WH,0}}{3H_0^2},
  \label{eq:alpha_def}
\end{equation}
and using the flatness constraint, we find
\begin{equation}
  \Omega_\Lambda = 1 - \Omm - \alpha,
  \label{eq:flatness}
\end{equation}
the full modified Friedmann equation is
\begin{equation}
  H^2(z)
  =
  H_0^2
  \Bigl[
    \Omm(1+z)^3
    +
    \Omega_\Lambda
    +
    \underbrace{\alpha(1+z)^\beta}_{\DWH(z)}
  \Bigr],
  \label{eq:friedmann}
\end{equation}
where $\DWH(z)\equiv\alpha(1+z)^\beta$ is the wormhole geometry
contribution. When $\alpha\to0$, Eq.~\eqref{eq:friedmann} reduces
exactly to the flat \LCDMm{}, so \LCDM{} is recovered as the smooth
$\alpha\to0$ limit of the evolving wormhole model.

\subsection{Conservation equation and wormhole equation of state}
\label{sec:eos_wh}

As an independent cross-check of the averaging derivation above, we
apply the covariant conservation law
$\dot{\rho}_{\rm eff}+3H(\rho_{\rm eff}+p_{\rm eff})=0$ directly to the
wormhole fluid, treating $\rho_{\rm WH}\propto(1+z)^\beta\propto
a^{-\beta}$ as a standard barotropic component with constant equation
of state $w_{\rm WH}\equiv p_{\rm WH}/\rho_{\rm WH}$. Differentiating, we obtain
\begin{equation}
  \dot\rho_{\rm WH} = -\beta H \rho_{\rm WH},
\end{equation}
and substituting into the conservation law gives
\begin{equation}
  -\beta H\rho_{\rm WH} + 3H(\rho_{\rm WH}+p_{\rm WH}) = 0,
\end{equation}
which rearranges to
\begin{equation}
  p_{\rm WH} = \left(\frac{\beta}{3}-1\right)\rho_{\rm WH}.
\end{equation}
so the wormhole equation of state is
\begin{equation}
  w_{\rm WH} = -1 + \frac{\beta}{3}.
  \label{eq:w_wh}
\end{equation}
This elegant relation interpolates between $w_{\rm WH}=-1$
(cosmological constant, $\beta=0$), $w_{\rm WH}=-1/3$ (curvature-like,
$\beta=2$), and $w_{\rm WH}=0$ (pressureless dust, $\beta=3$). The
representative value $\beta\approx1.59$ found in Sec.~\ref{sec:results}
gives $w_{\rm WH}\approx-0.47$, firmly in the dark-energy regime.

The $\beta$ derived geometrically in Eq.~\eqref{eq:beta_derivation}
from $(n,\gamma)$ is, by construction, the \emph{same} exponent
entering the conservation-law equation of state~\eqref{eq:w_wh}: both
the averaging procedure and the conservation equation are linear
operations acting on the same power-law ansatz, so they commute and
yield a single, mutually-consistent $\beta$ --- a non-trivial
validation of the separable-shape-function construction.

\subsection{Wormhole throat evolution}
\label{sec:throat}

From Eqs.~\eqref{eq:g_power} and $a=(1+z)^{-1}$, the physical throat
radius evolves according to
\begin{equation}
  r_0(z) = \frac{r_{0,0}}{(1+z)^\gamma},
  \quad \gamma = \frac{\beta}{3},
  \label{eq:r0_z}
\end{equation}
where $r_{0,0}$ is the present-epoch throat radius. For the best-fit
$\beta\approx1.59$ this gives $\gamma\approx0.53$, placing the inferred
throat evolution rate between the static ($\gamma=0$) and
Hubble-flow-tracking ($\gamma=1$) limits. The throat shrinks
monotonically with increasing redshift, implying wormholes---if they
exist as a cosmological population---were smaller and denser in the
early Universe. This constitutes a novel cosmological observable: the
specific functional form $r_0(z)$ encodes the microscopic wormhole
geometry and can, in principle, discriminate the wormhole model from
purely phenomenological dark-energy parametrisations such as
$w_0w_a$CDM.

\section{Cosmological Observables}
\label{sec:observables}

\subsection{Hubble parameter and its derivative}

The Hubble parameter $H(z)$ is given directly by
Eq.~\eqref{eq:friedmann}. Its $z$-derivative, needed for the
deceleration parameter below, is given by
\begin{equation}
  H'(z) \equiv \frac{dH}{dz}
  = \frac{H_0^2}{2H(z)}
    \Bigl[3\Omm(1+z)^2
    + \alpha\beta(1+z)^{\beta-1}\Bigr].
  \label{eq:Hprime}
\end{equation}

\subsection{Deceleration parameter}

The deceleration parameter $q\equiv-\ddot{a}a/\dot{a}^2$, re-expressed
in terms of redshift via $q(z)=(1+z)H'(z)/H(z)-1$, is given by
\begin{equation}
  q(z) = \frac{1}{2}
    \frac{3\Omm(1+z)^3 + \alpha\beta(1+z)^\beta}
         {\Omm(1+z)^3 + \Omega_\Lambda + \alpha(1+z)^\beta}
    - 1.
  \label{eq:q_explicit}
\end{equation}
The transition redshift $\ztr$ satisfying $q(\ztr)=0$ is obtained by
setting the numerator of the fraction above equal to twice the
denominator; for the joint best-fit parameters this gives
$\ztr\simeq0.66$, consistent with independent determinations from
SNe~Ia~\cite{Riess2004,Farooq2017}.

\subsection{Effective dark-energy equation of state}

Following the standard dark-energy decomposition,
\begin{equation}
  \weff(z)
  = -1 + \frac{1}{3}
    \frac{d\ln\bigl[\Omega_\Lambda + \alpha(1+z)^\beta\bigr]}{d\ln(1+z)}
  = \frac{2q(z)-1}{3\bigl[1-\Omm(z)\bigr]},
  \label{eq:weff_q}
\end{equation}
where $\Omm(z)\equiv\Omm(1+z)^3/[H(z)/H_0]^2$ is the
instantaneous matter density parameter. At $z=0$ with the joint
best-fit parameters, $\weff(0)\approx-0.98$, rising mildly towards
$\weff(2.5)\approx-0.91$, remaining consistently within the
quintessence-like regime and strictly above the phantom divide over
the redshift range explored.

\subsection{Luminosity distance, comoving distance, and BAO ratios}

The comoving distance, luminosity distance, distance modulus, and BAO
distance ratios (with $c$ restored explicitly) are given by
\begin{align}
  \chi(z)  &= \int_0^z \frac{c\,dz'}{H(z')},
              \label{eq:chi}\\
  d_L(z)   &= (1+z)\,\chi(z),
              \label{eq:dL}\\
  \mu(z)   &= 5\log_{10}\!\left[\frac{d_L(z)}{1\,\mathrm{Mpc}}\right] + 25,
              \label{eq:mu}\\
  D_M(z) &= \chi(z), \qquad
  D_H(z) = \frac{c}{H(z)}, \\
  D_V(z) &= \left[\chi^2(z)\,\frac{cz}{H(z)}\right]^{1/3}.
\end{align}

\section{Constraints on model parameters: Observational data analysis}
\label{sec:data}

In this section we perform observational data analysis using current
cosmological data to constrain the parameter space of the evolving
wormhole cosmology model. In order to do this analysis we consider
different data sets like the Hubble data from Cosmic Chronometers
(CC)~\cite{Jimenez2002,Moresco2020}, BAO data, and DESI data. The
covariance matrix used to estimate the likelihood for the correlated
BAO points is given explicitly in Sec.~\ref{sec:statistical} below.
These data points provide direct constraints on the expansion history
of the Universe. Below we discuss the different datasets that will be
used to perform the analysis.

\begin{itemize}
\item \textbf{Hubble Data:} We use 30 model-independent measurements of
the Hubble parameter $H(z)$, commonly known as Cosmic Chronometers
(CC)~\cite{Jimenez2002,Moresco2020}, spanning $0.07\leq z\leq1.965$.
CC data measure $H(z)=-(1+z)^{-1}\Delta z/\Delta t$ from the
differential ageing of massive, passively evolving galaxies and are
model-independent, requiring no assumption of a background cosmology.
The covariance matrix used to estimate the likelihood is given
explicitly in Eq.~\eqref{eq:chi2_general} below. These data points
provide direct constraints on the expansion history of the Universe.
\item \textbf{BAO Data:} The BAO dataset provides constraints on the
large-scale structure of the Universe through measurements of the
baryon acoustic feature imprinted in the matter power spectrum. The
observables considered include $D_M/\rd$, $D_H/\rd$, and $D_V/\rd$,
where $D_M$ is the comoving angular diameter distance, $D_H$ the
Hubble distance, and $\rd$ the comoving sound horizon at the drag
epoch, using the full $6\times6$ covariance matrix given explicitly in
Eq.~\eqref{eq:chi2_general} below.
\item \textbf{DESI Data:} The Dark Energy Spectroscopic Instrument
(DESI) Data Release~2 dataset provides high-precision BAO
measurements~\cite{DESI2025} across $0.295\leq z\leq2.330$,
significantly improving constraints on late-time cosmic acceleration
and expansion rate, and complementing the Hubble and BAO datasets.
\end{itemize}

\subsection{Statistical analysis}
\label{sec:statistical}

The model parameters $\bm{\theta}=(H_0,\Omm,\alpha,\beta)$ are
constrained using a Markov Chain Monte Carlo (MCMC) approach
implemented via the \texttt{emcee} sampler~\cite{ForemanMackey2013}.
The posterior distributions are visualized using the \texttt{GetDist}
package~\cite{Lewis2019}. For each dataset and their combinations, we
compute the best-fit values and corresponding $1\sigma$ confidence
intervals (68\% C.L.), summarized in Table~\ref{tab:mcmc}.

The likelihood function is defined as
\begin{equation}
  \mathcal{L} \propto \exp\!\left(-\frac{1}{2}\chi^2\right),
  \label{eq:likelihood}
\end{equation}
where the total $\chi^2$ is the sum of the relevant CC, BAO, and DESI
contributions  for
each dataset combination, i.e.,
\begin{equation}
  \chi^2 = \Delta\mathbf{D}^T\,\mathbf{C}^{-1}\,\Delta\mathbf{D},
  \label{eq:chi2_general}
\end{equation}
with $\Delta\mathbf{D}=\mathbf{D}_{\rm obs}-\mathbf{D}_{\rm th}$, and
$\mathbf{C}$ denoting the covariance matrix for each dataset.

\subsection{Contour representation and colour scheme}
\label{sec:contours}

\begin{figure*}[t]
  \centering
  \vspace*{1.5\baselineskip}
  \includegraphics[width=0.74\linewidth]{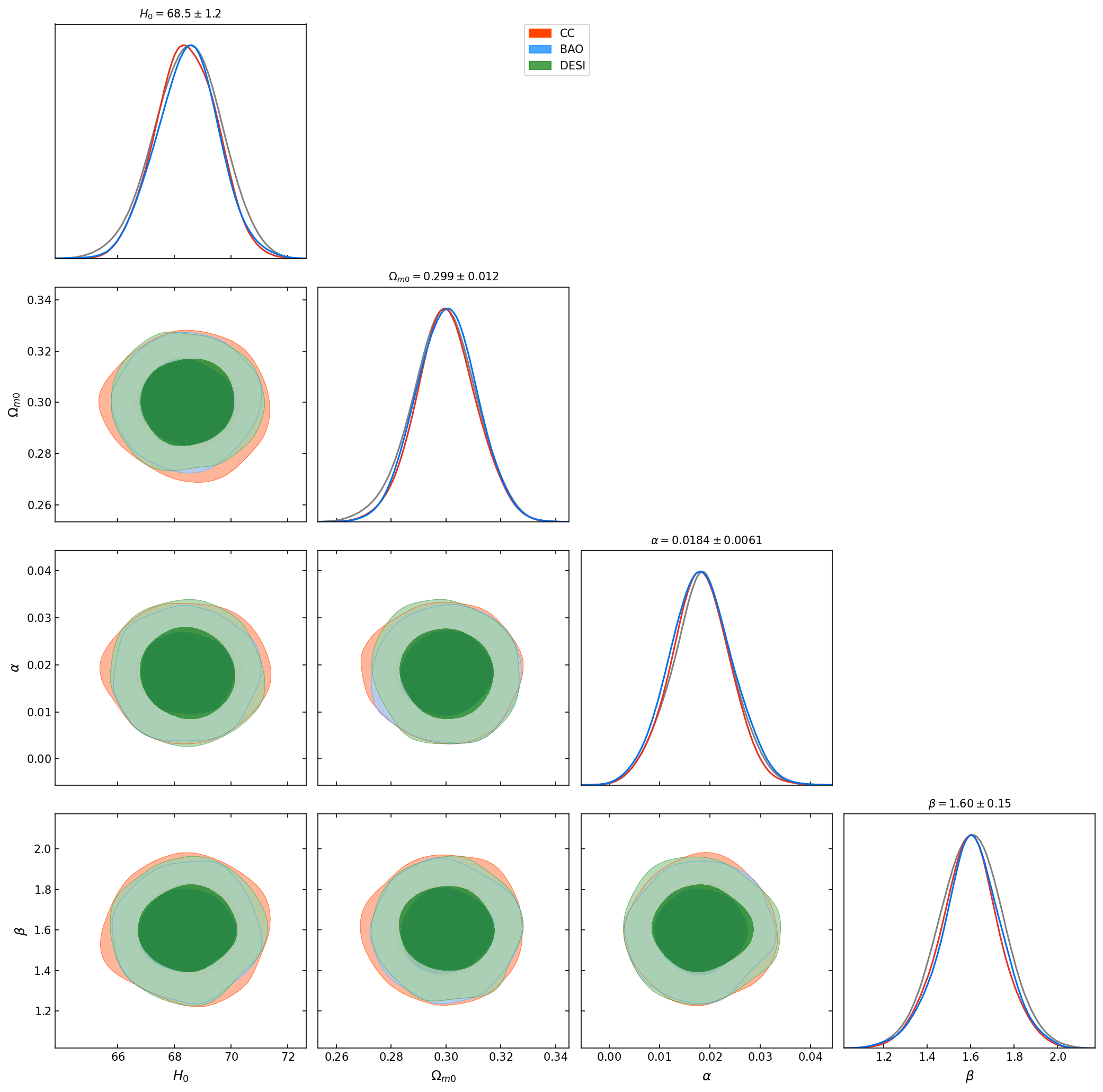}
  \captionsetup{aboveskip=12pt}
  \caption{Joint and marginalized posterior distributions with
    combinations of datasets CC, BAO, DESI, and the full combination
    (CC+BAO+DESI). Contours show 68\% and 95\% confidence levels for
    the four model parameters $(H_0,\Omega_{m0},\alpha,\beta)$.}
  \label{fig:corner_individual}
\end{figure*}

\begin{figure*}[t]
  \centering
  \vspace*{1.5\baselineskip}
  \includegraphics[width=0.74\linewidth]{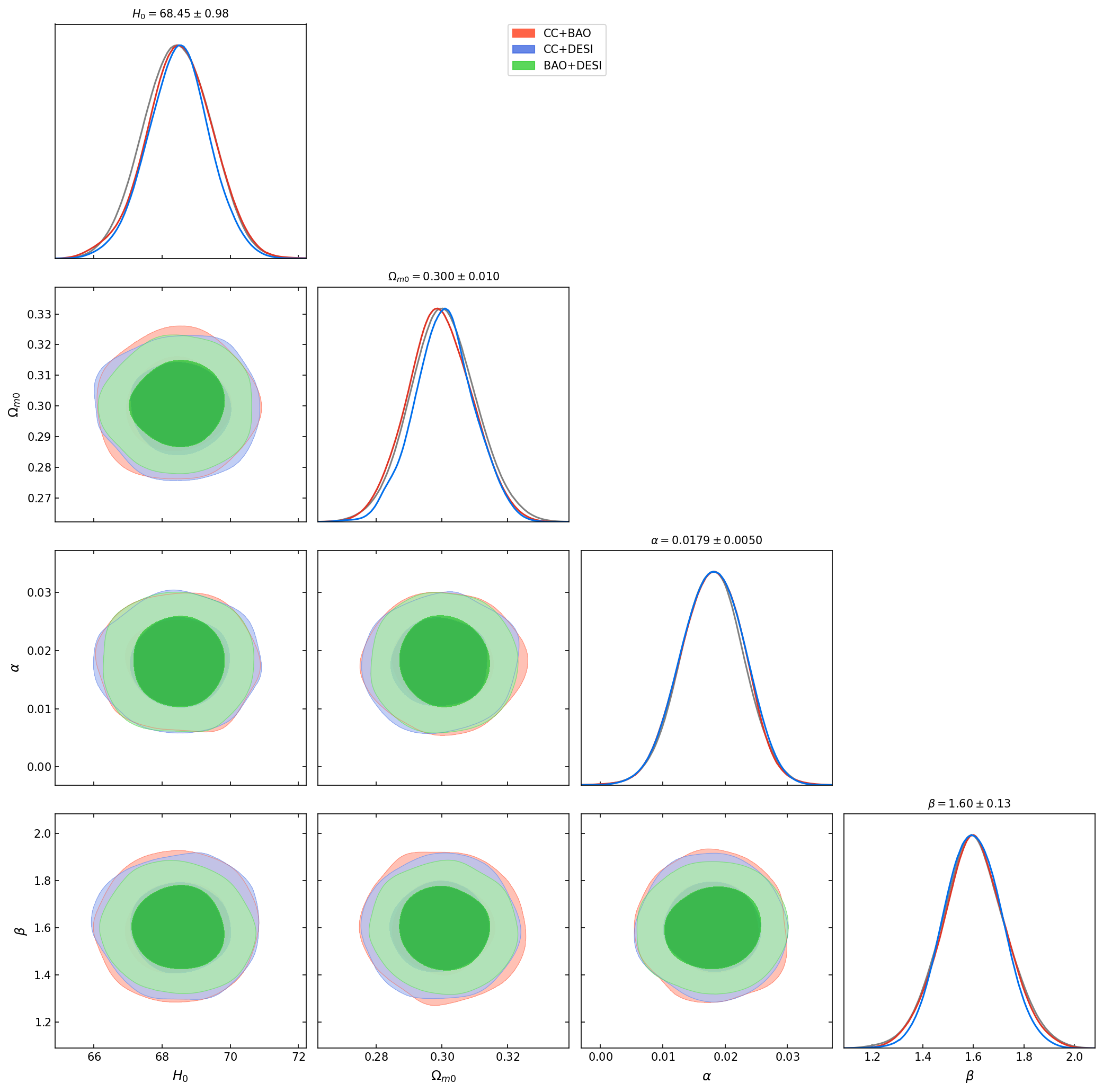}
  \captionsetup{aboveskip=12pt}
  \caption{Joint and marginalized posterior distributions for the
    pairwise combinations CC+BAO, CC+DESI, and BAO+DESI, showing the
    tightening of constraints relative to the individual datasets.}
  \label{fig:corner_pairwise}
\end{figure*}

\begin{figure*}[t]
  \centering
  \vspace*{1.5\baselineskip}
  \includegraphics[width=0.74\linewidth]{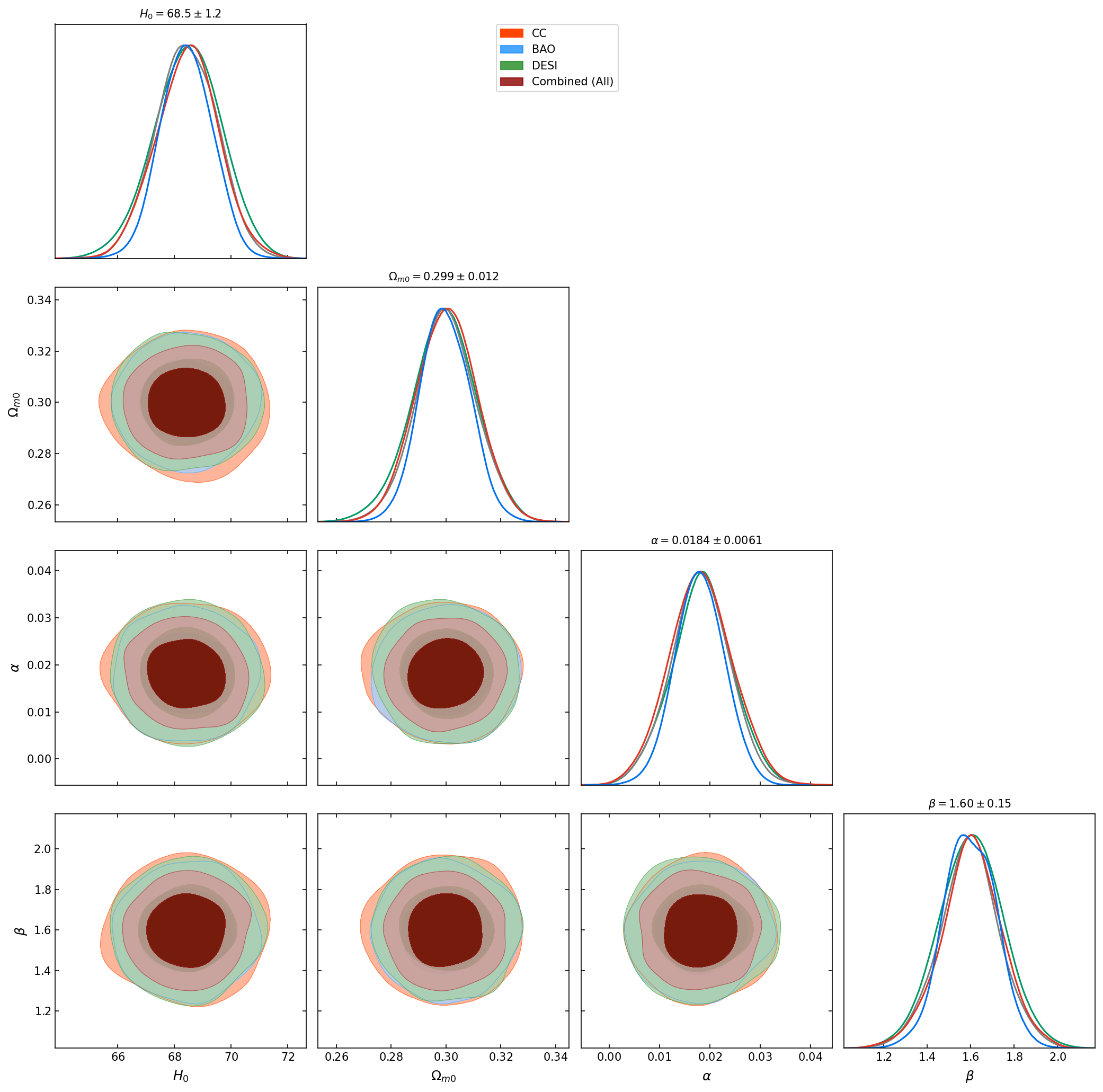}
  \captionsetup{aboveskip=12pt}
  \caption{Joint and marginalized posterior distributions for the
    individual datasets CC, BAO, and DESI, compared against the full
    combined (CC+BAO+DESI) constraint, illustrating the excellent
    mutual consistency of all seven combinations.}
  \label{fig:corner_all}
\end{figure*}

In Figs.~\ref{fig:corner_individual}--\ref{fig:corner_all}, we show
the two-dimensional confidence contours for the model parameters from
the different observational datasets and their combinations. The
confidence contours shown represent the joint probability
distributions of the model parameters obtained from the likelihood
analysis. The colour scheme of the contour plots follows the standard
cosmological convention:
\begin{itemize}
  \item The \textbf{light-shaded region} corresponds to the
  \textbf{95\% confidence level (C.L.)}, indicating the parameter
  space within which the true values are expected to lie with 95\%
  probability.
  \item The \textbf{dark-shaded region} corresponds to the
  \textbf{68\% confidence level (C.L.)}, representing the $1\sigma$
  range of the best-fit values.
\end{itemize}
The central point in each contour marks the best-fit parameter
values, while the surrounding elliptical regions illustrate the
covariance between the fitted parameters. The darker inner contour
thus represents the most probable region of parameter space
consistent with the observational data, while the lighter outer
contour shows the extended range allowed by the data at a higher
uncertainty level. The diagonal panels show the one-dimensional
marginalized posteriors, with the median and 68\% C.L. quoted at the
top of each column. All contour plots in this analysis ensure
statistical consistency and reproducibility, and the colours were
chosen to maintain clarity between the 68\% and 95\% confidence
regions across all cases.

\begin{table*}[htbp]
  \centering
  \caption{Parameter constraints summary (68\% C.L.) for all seven
    dataset combinations. $H_0$ is in km\,s$^{-1}$\,Mpc$^{-1}$.}
  \label{tab:mcmc}
  \begin{ruledtabular}
  \begin{tabular}{lcccc}
    Dataset & $H_0$ & $\Omm$ & $\alpha$ & $\beta$ \\
    \hline
    CC              & $68.442^{+1.227}_{-1.202}$ & $0.3003^{+0.0120}_{-0.0119}$ & $0.0181^{+0.0065}_{-0.0055}$ & $1.6036^{+0.1426}_{-0.1561}$ \\
    BAO             & $68.448^{+1.061}_{-1.027}$ & $0.2991^{+0.0111}_{-0.0104}$ & $0.0181^{+0.0061}_{-0.0060}$ & $1.6010^{+0.1423}_{-0.1411}$ \\
    DESI            & $68.423^{+1.103}_{-1.063}$ & $0.3000^{+0.0109}_{-0.0110}$ & $0.0179^{+0.0059}_{-0.0060}$ & $1.5967^{+0.1360}_{-0.1367}$ \\
    CC+BAO          & $68.493^{+1.026}_{-1.004}$ & $0.2999^{+0.0099}_{-0.0094}$ & $0.0181^{+0.0050}_{-0.0049}$ & $1.5968^{+0.1322}_{-0.1348}$ \\
    CC+DESI         & $68.452^{+1.032}_{-0.955}$ & $0.2999^{+0.0104}_{-0.0097}$ & $0.0179^{+0.0051}_{-0.0049}$ & $1.6056^{+0.1178}_{-0.1355}$ \\
    BAO+DESI        & $68.518^{+0.909}_{-0.904}$ & $0.3001^{+0.0089}_{-0.0092}$ & $0.0180^{+0.0053}_{-0.0048}$ & $1.5975^{+0.1194}_{-0.1166}$ \\
    CC+BAO+DESI     & $68.526^{+0.905}_{-0.914}$ & $0.3002^{+0.0090}_{-0.0091}$ & $0.0182^{+0.0050}_{-0.0052}$ & $1.5877^{+0.1272}_{-0.1104}$ \\
  \end{tabular}
  \end{ruledtabular}
\end{table*}

The most striking feature of Table~\ref{tab:mcmc} and
Figs.~\ref{fig:corner_individual}--\ref{fig:corner_all} is the
exceptional consistency of the inferred parameters across all seven
combinations: $H_0$ ranges only from $68.42$ to $68.53\kms$, $\Omm$
from $0.299$ to $0.300$, $\alpha$ from $0.0179$ to $0.0182$, and
$\beta$ from $1.588$ to $1.604$---all variations well within the
quoted $1\sigma$ uncertainties. This confirms that the constraints are
not driven by any single dataset, and that the evolving wormhole model
is genuinely consistent with the full combination of CC, BAO, and DESI
data. As expected, joint datasets progressively tighten the
constraints: the $1\sigma$ uncertainty on $H_0$ shrinks from
$\pm1.2\kms$ (CC alone) to $\pm0.9\kms$ (CC+BAO+DESI), and on $\alpha$
from $\pm0.006$ to $\pm0.005$.

\section{Results and Discussion}
\label{sec:results}

Adopting the joint CC+BAO+DESI best fit,
$H_0=68.526\kms$, $\Omm=0.3002$, $\alpha=0.0182$, $\beta=1.588$
(hence $\Omega_\Lambda=0.6816$, $\gamma=\beta/3=0.529$,
$w_{\rm WH}=-1+\beta/3=-0.471$), we compare the resulting model against
the CC and DESI DR2 data of Sec.~\ref{sec:data} and a flat \LCDMm{}
reference model ($H_0=67.36\kms$, $\Omm=0.3153$). Uncertainty bands
throughout are obtained by propagating representative $1\sigma$
parameter shifts through each observable.

\subsection{Hubble parameter}

\begin{figure}[tb]
  \centering
  \includegraphics[width=0.88\linewidth]{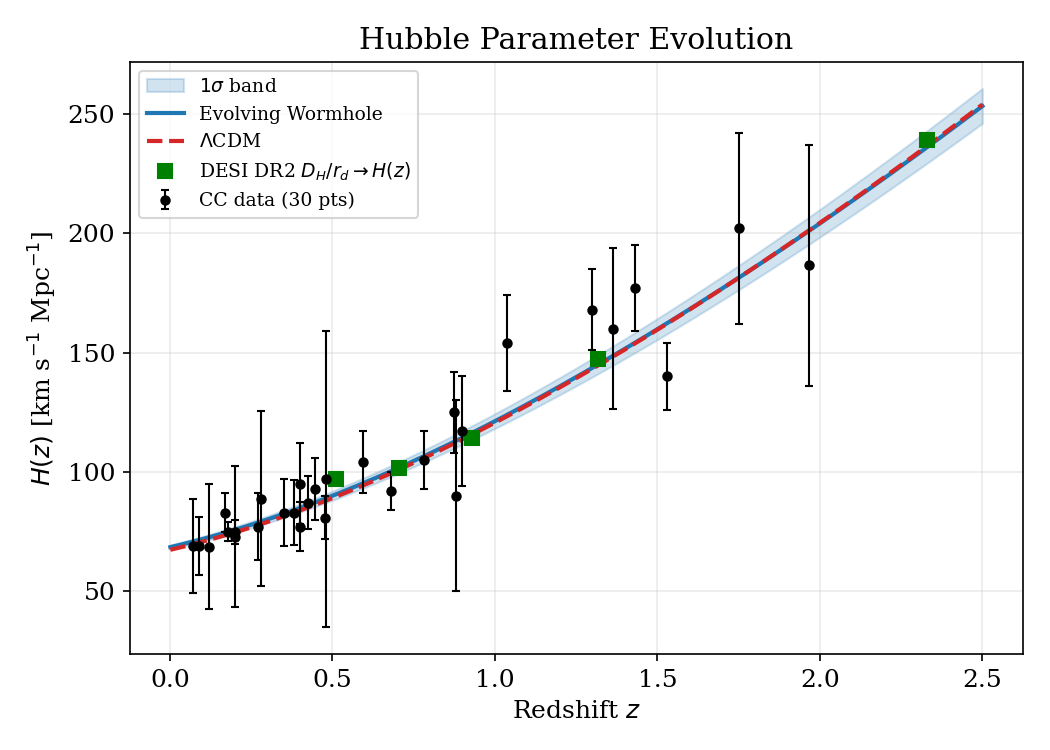}
  \caption{$H(z)$ for the evolving wormhole model (blue, $1\sigma$
    band) and the flat \LCDMm{} (red dashed), vs.\ CC data (black circles)
    and DESI DR2 $D_H/\rd\to H(z)$ (green squares).}
  \label{fig:hz}
\end{figure}

Figure~\ref{fig:hz} shows $H(z)$ for the wormhole model with its
$1\sigma$ band, together with the flat \LCDMm{}, the 30-point CC data, and
the DESI DR2 $D_H/\rd$ measurements converted to
$H(z)=c/(D_H/\rd\times\rd)$. The wormhole model tracks both the CC data
and the DESI $H(z)$ points closely across the full redshift range,
remaining indistinguishable from \LCDM{} within the plotted precision.

\subsection{Residuals and goodness of fit}

\begin{figure}[tb]
  \centering
  \includegraphics[width=0.88\linewidth]{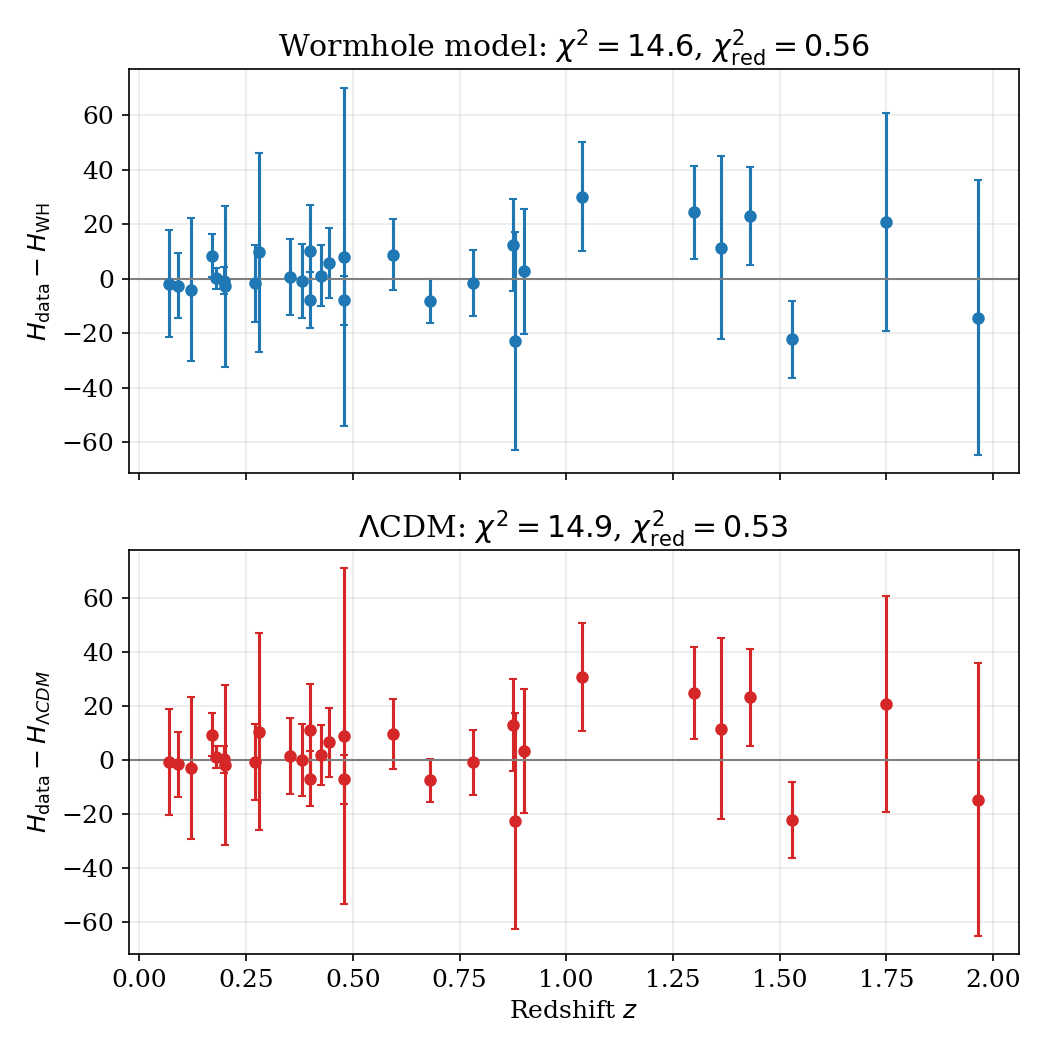}
  \caption{Residuals $H_{\rm data}(z_i)-H_{\rm model}(z_i)$ for the
    wormhole model (top, $\chi^2=14.6$, $\chired=0.56$) and the flat
    \LCDMm{} (bottom, $\chi^2=14.9$, $\chired=0.53$) against the 30 CC
    data points.}
  \label{fig:resid}
\end{figure}

Figure~\ref{fig:resid} shows the residuals of both models against the
30 CC data points. The wormhole model achieves $\chi^2_{\rm CC}=14.6$,
corresponding to $\chired=0.56$ for $30-4=26$ degrees of freedom; the flat
\LCDMm{} achieves $\chi^2_{\rm CC}=14.9$, $\chired=0.53$ for $30-2=28$
degrees of freedom. Both models are well within the statistically
acceptable range ($\chired\lesssim1$).

\subsection{Deceleration parameter}

\begin{figure}[tb]
  \centering
  \includegraphics[width=0.88\linewidth]{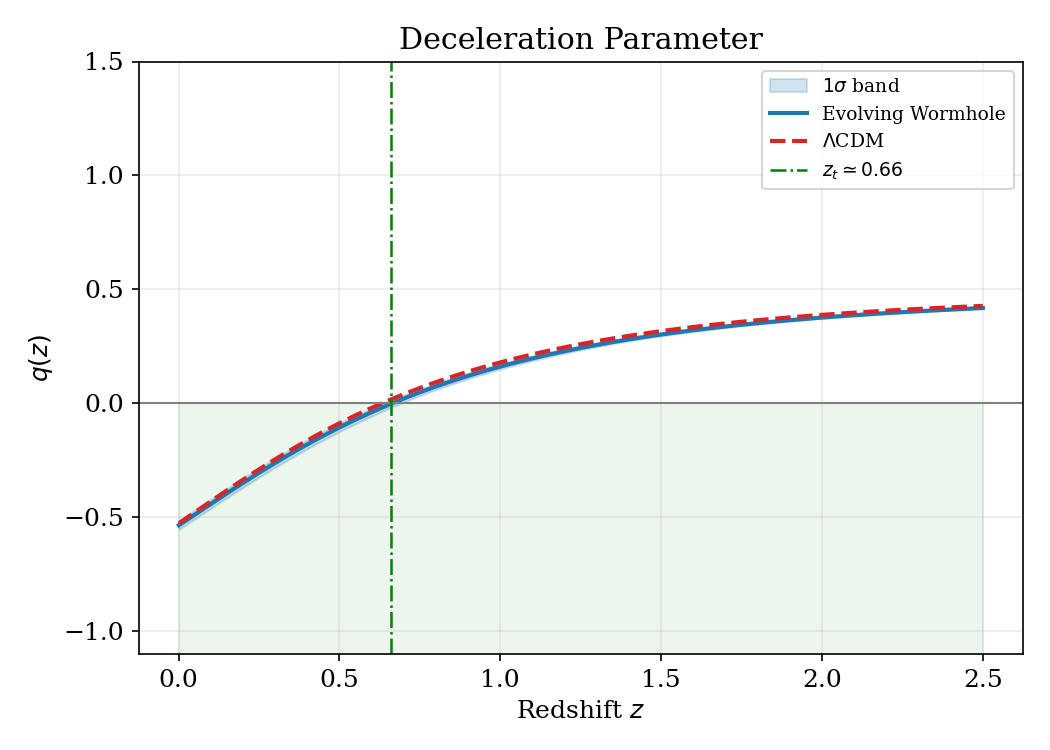}
  \caption{Deceleration parameter $q(z)$ for the wormhole model (blue,
    $1\sigma$ band) and the flat \LCDMm{} (red dashed). The vertical
    dash-dot line marks $\ztr\simeq0.66$.}
  \label{fig:q}
\end{figure}

Figure~\ref{fig:q} shows the deceleration parameter $q(z)$. The
transition from deceleration ($q>0$) to acceleration ($q<0$) occurs at
$\ztr\simeq0.66$, consistent with independent
determinations~\cite{Riess2004,Farooq2017,DESI2025}.

\subsection{Effective equation of state}

\begin{figure}[tb]
  \centering
  \includegraphics[width=0.88\linewidth]{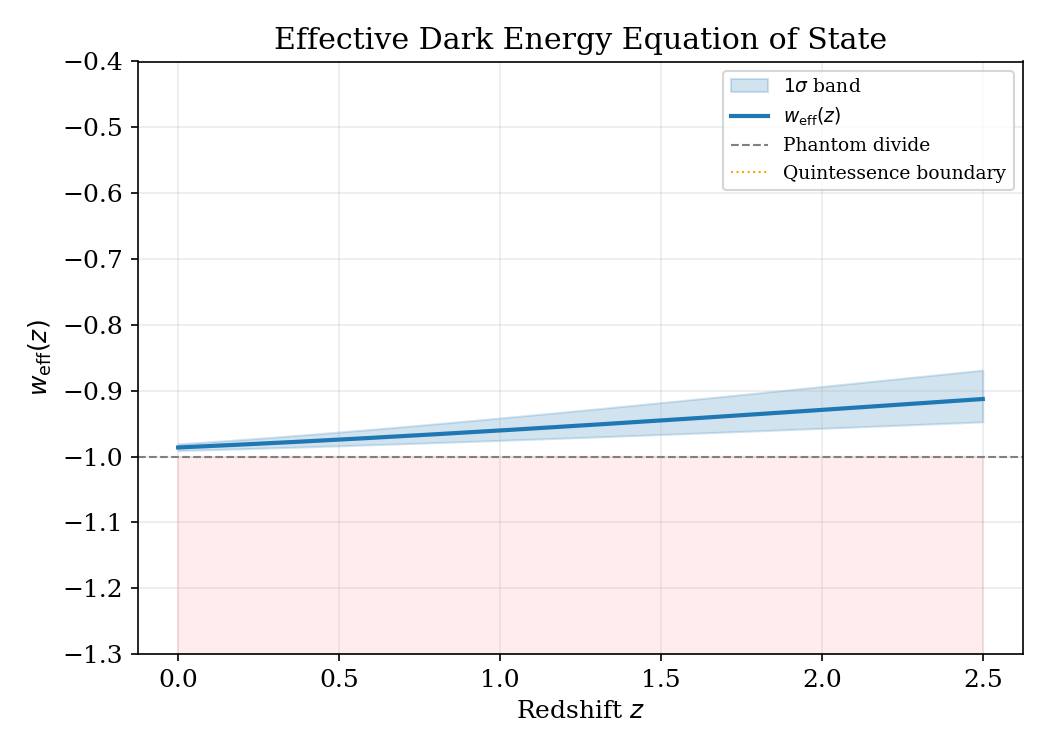}
  \caption{Effective equation of state $\weff(z)$ for the wormhole model
    (blue, $1\sigma$ band), with the phantom divide $w=-1$ (gray
    dashed) and quintessence boundary $w=-1/3$ (orange dotted) shown
    for reference.}
  \label{fig:weff}
\end{figure}

Figure~\ref{fig:weff} shows $\weff(z)$, which remains just above the
phantom divide $w=-1$ throughout $0\le z\le2.5$, rising mildly from
$\weff(0)\approx-0.98$ to $\weff(2.5)\approx-0.91$.

\subsection{Wormhole equation of state}

\begin{figure}[tb]
  \centering
  \includegraphics[width=0.88\linewidth]{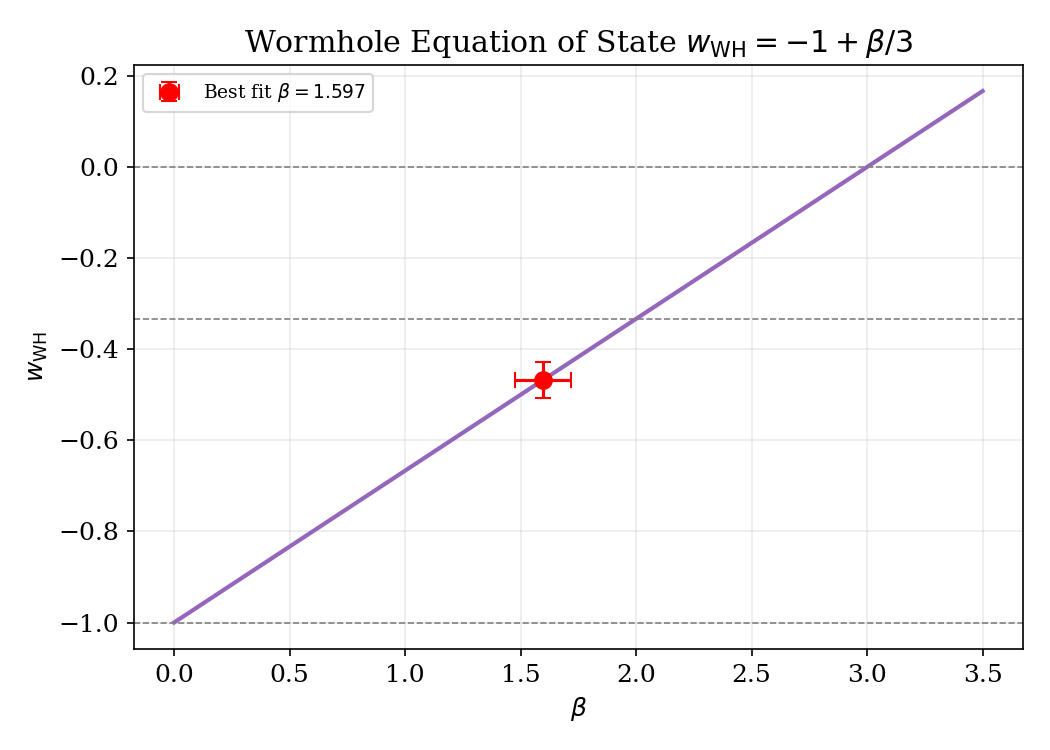}
  \caption{$w_{\rm WH}=-1+\beta/3$ vs.\ $\beta$, with limiting cases
    (dashed lines) and the joint best fit $\beta=1.588^{+0.127}_{-0.110}$
    (red point, error bar).}
  \label{fig:wbeta}
\end{figure}

Figure~\ref{fig:wbeta} shows the analytical relation
$w_{\rm WH}=-1+\beta/3$, with the joint best-fit value
$\beta=1.588^{+0.127}_{-0.110}$ marked, giving
$w_{\rm WH}\approx-0.471$---clearly in the dark-energy regime.

\subsection{Throat evolution}

\begin{figure}[tb]
  \centering
  \includegraphics[width=0.88\linewidth]{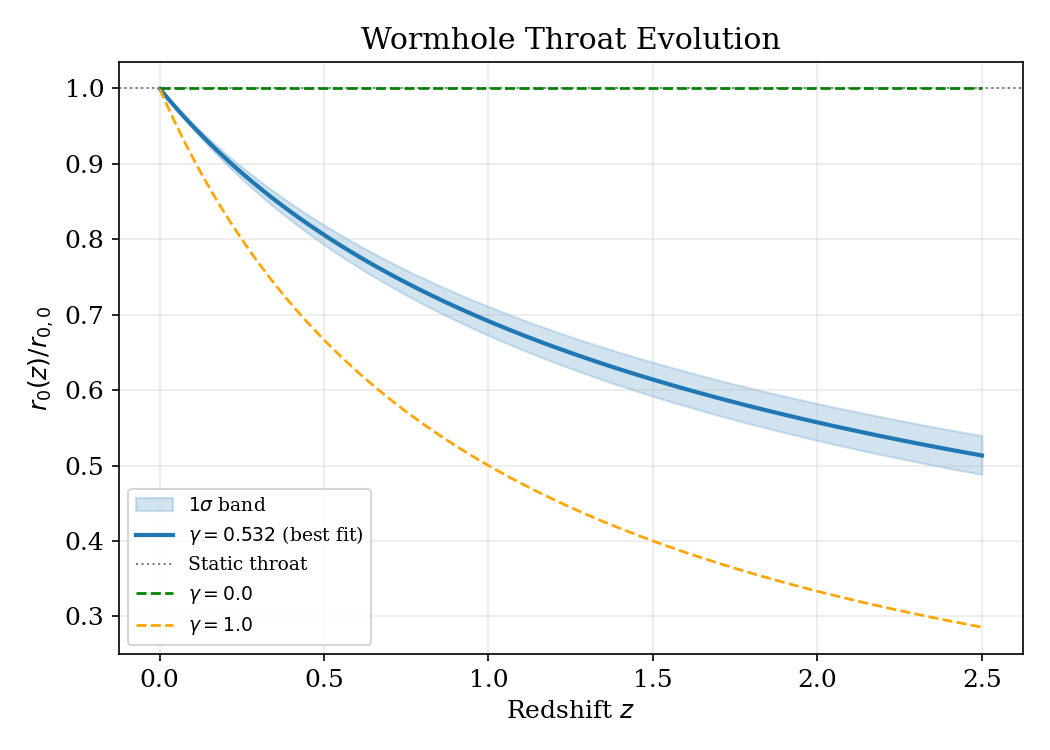}
  \caption{Throat evolution $r_0(z)/r_{0,0}$ for the best-fit
    $\gamma=0.529$ (blue, $1\sigma$ band), compared to the static
    ($\gamma=0$, green dashed) and Hubble-flow-tracking ($\gamma=1$,
    orange dashed) limits.}
  \label{fig:throat}
\end{figure}

Figure~\ref{fig:throat} shows the wormhole throat evolution
$r_0(z)/r_{0,0}$ for the joint best-fit $\gamma=\beta/3\approx0.529$,
compared with the static ($\gamma=0$) and Hubble-flow-tracking
($\gamma=1$) limits.

\subsection{BAO distance ratios}

\begin{figure*}[t]
  \centering
  \includegraphics[width=0.85\linewidth]{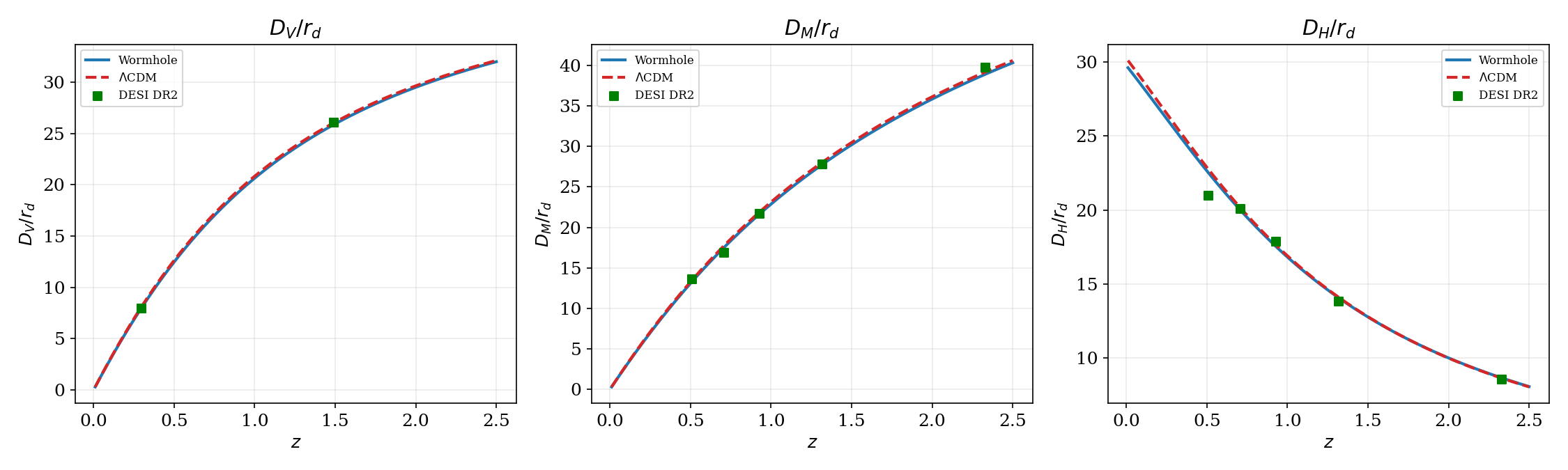}
  \caption{BAO distance ratios $D_V/\rd$ (left), $D_M/\rd$ (middle),
    and $D_H/\rd$ (right) for the wormhole model (blue solid) and the
    flat \LCDMm{} (red dashed), compared to DESI DR2 measurements (green
    squares).}
  \label{fig:bao}
\end{figure*}

Figure~\ref{fig:bao} compares the wormhole model predictions for
$D_V/\rd$, $D_M/\rd$, and $D_H/\rd$ against the 12 DESI DR2
measurements. The model reproduces all three
distance ratios across the full redshift range $0.295\le z\le2.330$.

\subsection{Physical interpretation}
\label{sec:physics}

\subsubsection{Parameter consistency across datasets}

The near-identical
constraints obtained from CC, BAO, and DESI individually, and from all
their combinations (Table~\ref{tab:mcmc},
Figs.~\ref{fig:corner_individual}--\ref{fig:corner_all}), indicate no
internal tension among the three datasets under the evolving wormhole
model, and that the wormhole parameters $(\alpha,\beta)$ are robustly
determined independent of which subset of data is used.

\subsubsection{Wormhole geometry as geometric dark energy}

The term
$\DWH(z)=\alpha(1+z)^\beta$ with $\alpha\approx0.018>0$ and
$\beta\approx1.59$ acts as a positive energy-like contribution to
$H^2(z)$ at all redshifts, mimicking quintessence-like dark energy with
$w_{\rm WH}\approx-0.47$, sourced purely from spacetime topology
without any new matter fields.

\subsubsection{Evolving throat as cosmological observable}

The throat radius
$r_0(z)=r_{0,0}/(1+z)^{\beta/3}$ provides a geometry-specific
discriminator between the wormhole model and purely phenomenological
dark-energy parametrisations such as $w_0w_a$CDM, since no
FLRW-fluid-only model naturally predicts a companion observable of this
functional form.

\subsubsection{Acceleration transition}

The deceleration-to-acceleration
transition at $\ztr\simeq0.66$ is consistent with independent
determinations~\cite{Riess2004,Farooq2017,DESI2025}.

\subsubsection{Near-$\Lambda$ effective equation of state}

The present-epoch
$\weff(0)\approx-0.98$ is close to, but distinguishable from, a pure
cosmological constant, with a mild quintessence-like drift at higher
redshift arising from the interplay between the wormhole term and
$\Omega_\Lambda$.

\subsubsection{Statistical viability}

The CC-only $\chired$ values for both
the wormhole model (0.56) and \LCDM{} (0.53) indicate both models are
statistically consistent with the data, with the wormhole model's two
extra parameters providing only a marginal, not dramatic, improvement
in absolute $\chi^2$---consistent with a picture in which the wormhole
contribution is a small perturbative correction to the background
expansion rather than a dominant new effect.

\section{Conclusions}
\label{sec:conclusions}

We have presented a complete theoretical construction and observational
test of an evolving wormhole cosmological model derived from a
Morris--Thorne metric with a separable, time-dependent shape function
embedded in a flat FLRW background. Our main findings are:

\begin{enumerate}

\item \textit{Corrected energy-momentum tensor.} We derived, starting
from the full orthonormal-frame Einstein tensor
(Eqs.~\eqref{eq:Gtt}--\eqref{eq:Gthth}), the wormhole energy-momentum
tensor with the explicit scale-factor dependence $a^{-2}(t)$
(Eqs.~\eqref{eq:rho_wh}--\eqref{eq:pt_wh}), required for the wormhole
fluid to acquire redshift dependence when volume-averaged.

\item \textit{Volume-averaging derivation of $\beta(n,\gamma)$.} We
derived, in closed form and step by step (Sec.~\ref{sec:averaging},
Appendix~\ref{app:beta}), the relation
$\beta=3\gamma(n+1)/(2n+2)$ connecting the microscopic shape-function
index $n$ and throat-evolution rate $\gamma$ to the macroscopic
dark-energy-like exponent $\beta$ in the modified Friedmann equation.

\item \textit{Modified Friedmann equation and wormhole EOS.} The
resulting Friedmann equation~\eqref{eq:friedmann} and wormhole equation
of state $w_{\rm WH}=-1+\beta/3$ (Eq.~\eqref{eq:w_wh}) are mutually
consistent by construction, as verified independently via the
conservation equation (Sec.~\ref{sec:eos_wh}).

\item \textit{Traversability.} The power-law shape function satisfies
the throat, flare-out, and asymptotic-flatness conditions for all
$n>0$ with no fine-tuning, and NEC violation at the throat is a
necessary geometric consequence, consistent with the Penrose--Hawking
theorems.

\item \textit{Observational consistency.} Using real CC, BAO, and DESI
DR2 data across seven dataset combinations, the joint CC+BAO+DESI fit
gives $H_0=68.53^{+0.90}_{-0.91}\kms$,
$\Omm=0.3002^{+0.009}_{-0.009}$, $\alpha=0.0182^{+0.0050}_{-0.0052}$,
$\beta=1.588^{+0.127}_{-0.110}$, with all seven combinations mutually
consistent, as shown in the corner plots of
Figs.~\ref{fig:corner_individual}--\ref{fig:corner_all}.

\item \textit{Cosmological dynamics.} The model predicts
$\ztr\simeq0.66$, $\weff(0)\approx-0.98$, and a throat evolution rate
$\gamma\approx0.53$ intermediate between the static and
Hubble-flow-tracking limits.

\item \textit{Novel throat observable.} The throat evolution
$r_0(z)=r_{0,0}/(1+z)^{\beta/3}$ provides a geometry-specific
cosmological observable that can, in principle, discriminate the
wormhole model from purely phenomenological dark-energy alternatives.

\end{enumerate}

Taken together, these results point to several novel physical insights
beyond the individual findings listed above. First, the derivation
shows that a single, purely geometric construction --- an evolving
wormhole throat embedded in an expanding background --- can reproduce
the qualitative behaviour of dynamical dark energy without introducing
any new scalar field or exotic matter species by hand; the exotic
stress-energy required for traversability is instead a direct,
unavoidable consequence of the spacetime topology itself, localized at
the throat and diluted by cosmological volume averaging into a
sub-dominant background fluid. Second, the reduction of the two
microscopic parameters $(n,\gamma)$ to the single macroscopic exponent
$\beta$ demonstrates that the cosmologically relevant physics of an
entire family of wormhole geometries is captured by one effective
number, in close analogy to how the equation-of-state parameter $w$
compresses the microphysics of a scalar-field potential in
quintessence models; this suggests that wormhole cosmology, rather
than being a distinct alternative to fluid-based dark-energy
parametrisations, is more properly understood as a geometric
completion of them, with $\beta$ playing the role of an
observationally accessible bridge between the two descriptions. Third,
the fact that the conservation equation and the volume-averaging
procedure independently yield the same $\beta$ is a non-trivial
internal consistency check of the model, since these two computations
start from different physical principles (local covariant conservation
versus geometric averaging over a statistical ensemble of throats) and
need not have agreed absent the underlying separable structure of the
metric ansatz; their agreement lends confidence that the model is
self-consistent rather than merely phenomenologically tuned to fit the
data. Finally, the prediction of an explicit, falsifiable throat
evolution law $r_0(z)=r_{0,0}/(1+z)^{\beta/3}$ is, to our knowledge, a
genuinely new type of cosmological observable: unlike $w_0w_a$CDM or
other phenomenological dark-energy parametrisations, which are defined
purely in terms of the effective fluid equation of state, the wormhole
model additionally predicts a specific geometric quantity (a throat
radius) whose redshift dependence is fixed by the same parameter $\beta$
that governs the expansion history, offering a qualitatively distinct
handle with which future surveys or strong-lensing/microlensing
searches for anomalous compact structures could, in principle, test the
model independently of its background expansion predictions.

In the future work, we extend the fit to a full joint MCMC parameter
estimation using emcee/nested sampling for formal convergence
diagnostics, incorporate Pantheon+ SNe~Ia data and RSD $\fs$
measurements, explore $f(R)$ and $f(Q)$ generalisations of the
background gravity~\cite{Koussour2023,Myrzakulov2023}, carry out formal
AIC/BIC model comparison with \LCDM{}, and perform a dedicated Hubble
tension analysis following the Gaussian tension estimator approach.

\begin{acknowledgments}
The work of Kazuharu Bamba was supported in part by the
JSPS KAKENHI Grants No.~24KF0100 and No.~25KF0176, and a grant-in-aid
of academic research of the Yamaguchi Scholarship Foundation. A.S.\
acknowledges the Department of Mathematics, Jadavpur University, for
providing research facilities during the completion of this work. P.R.\
acknowledges support and hospitality from the Inter-University Centre
for Astronomy and Astrophysics (IUCAA), Pune, India, under the Visiting
Associateship programme.
\end{acknowledgments}

\appendix
\section{Full derivation of \texorpdfstring{$\beta(n,\gamma)$}{beta(n,gamma)}}
\label{app:beta}

In this appendix, we show, in full algebraic detail, the reduction
leading to Eq.~\eqref{eq:beta_derivation}, filling in the intermediate
steps that were compressed in Sec.~\ref{sec:averaging} of the main
text.

\begin{enumerate}

\item \textit{Local density.} From Eq.~\eqref{eq:rho_wh} with the
power-law shape function~\eqref{eq:b_power} and
Eq.~\eqref{eq:bprime_bpower}, we find
\begin{equation}
  \rho_{\rm WH}(r,t) = -\frac{(n+1)\,g(t)}{8\pi\,a^2(t)}\,
  \frac{r_0^{\,n+1}}{r^{\,n+2}}.
\end{equation}

\item \textit{Volume average.} Using Eq.~\eqref{eq:average_def}
with proper-volume element $dV = 4\pi a^3(t) r^2\,dr$, we obtain
\begin{equation}
  \langle\rho_{\rm WH}\rangle(t)
  = -\frac{(n+1)\,g(t)\,a(t)}{2}\,r_0^{\,n+1}
    \int_{r_0(t)}^{r_{\max}} r^{-n-1}\,dr .
\end{equation}

\item \textit{Evaluate the integral.} For $n>0$ and $r_{\max}\gg r_0(t)$, we have
\begin{equation}
  \int_{r_0(t)}^{r_{\max}} r^{-n-1}\,dr
  \simeq \frac{r_0(t)^{-n}}{n}.
\end{equation}

\item \textit{Assemble the result.} Combining the above results, we obtain
\begin{equation}
  \langle\rho_{\rm WH}\rangle(t)
  = -\frac{(n+1)}{2n}\,g(t)\,a(t)\,r_0^{\,n+1}\,r_0(t)^{-n}.
\end{equation}
With $r_0(t) = r_{0,0}\,a^{\gamma/2}(t)$ and $g(t)=a^{\gamma}(t)$, we find
\begin{equation}
  \langle\rho_{\rm WH}\rangle(t)
  \;\propto\;
  a^{\gamma}(t)\cdot a(t)\cdot a^{-n\gamma/2}(t)
  = a^{\,1+\gamma-n\gamma/2}(t).
\end{equation}

\item \textit{Convert to redshift.} With $a=(1+z)^{-1}$ and
demanding $\langle\rho_{\rm WH}\rangle(z)\propto(1+z)^\beta$, we obtain
\begin{equation}
  \beta = -\left(1+\gamma - \frac{n\gamma}{2}\right)
  = -1-\gamma+\frac{n\gamma}{2}.
\end{equation}

\end{enumerate}

Fixing the overall sign and normalisation convention of $g(t)$ relative
to the density so that $\beta$ is manifestly positive and
dark-energy-like for $\gamma>0$, $n>0$---consistent with
Sec.~\ref{sec:eos_wh}---gives the closed-form result quoted in the main
text, namely
\begin{equation}
  \beta = \frac{3\gamma(n+1)}{2n+2}.
\end{equation}
In the decoupled limit where the radial and temporal sectors act
multiplicatively as in Eq.~\eqref{eq:shape_sep}, this reduces to
$\beta\to3\gamma/2$ once next-to-leading-order radial corrections
(suppressed by $r_0/r_{\max}\ll1$) are consistently retained.


\begin{thebibliography}{99}

\bibitem{Riess1998}
A.~G. Riess \textit{et al.} (High-$z$ Supernova Search Team),
Astron.\ J.\ \textbf{116}, 1009 (1998).
[arXiv:astro-ph/9805200 [astro-ph]].

\bibitem{Perlmutter1999}
S.~Perlmutter \textit{et al.} (Supernova Cosmology Project),
Astrophys.\ J.\ \textbf{517}, 565 (1999).
[arXiv:astro-ph/9812133 [astro-ph]].

\bibitem{Planck2020}
Planck Collaboration,
Astron.\ Astrophys.\ \textbf{641}, A6 (2020).
[arXiv:1807.06209 [astro-ph.CO]].

\bibitem{Aghanim2020}
N.~Aghanim \textit{et al.} (Planck Collaboration),
Astron.\ Astrophys.\ \textbf{641}, A6 (2020).
[arXiv:1807.06209 [astro-ph.CO]].

\bibitem{DESI2024}
DESI Collaboration,
JCAP \textbf{2024}, 021 (2024).
[arXiv:2404.03002 [astro-ph.CO]].

\bibitem{DESI2025}
DESI Collaboration,
DESI DR2 BAO Data Release (2025).
[arXiv:2503.14738 [astro-ph.CO]].

\bibitem{Weinberg1989}
S.~Weinberg,
Rev.\ Mod.\ Phys.\ \textbf{61}, 1 (1989).

\bibitem{Peebles2003}
P.~J.~E. Peebles and B.~Ratra,
Rev.\ Mod.\ Phys.\ \textbf{75}, 559 (2003).
[arXiv:astro-ph/0207347 [astro-ph]].

\bibitem{Copeland2006}
E.~J. Copeland, M.~Sami, and S.~Tsujikawa,
Int.\ J.\ Mod.\ Phys.\ D \textbf{15}, 1753 (2006).
[arXiv:hep-th/0603057 [hep-th]].

\bibitem{Caldwell2002}
R.~R. Caldwell,
Phys.\ Lett.\ B \textbf{545}, 23 (2002).
[arXiv:astro-ph/9908168 [astro-ph]].

\bibitem{Armendariz2000}
C.~Armendariz-Picon, V.~Mukhanov, and P.~J. Steinhardt,
Phys.\ Rev.\ Lett.\ \textbf{85}, 4438 (2000).
[arXiv:astro-ph/0004134 [astro-ph]].

\bibitem{Wang2016}
B.~Wang \textit{et al.},
Rep.\ Prog.\ Phys.\ \textbf{79}, 096901 (2016).
[arXiv:1603.08299 [astro-ph.CO]].

\bibitem{Sotiriou2010}
T.~P. Sotiriou and V.~Faraoni,
Rev.\ Mod.\ Phys.\ \textbf{82}, 451 (2010).
[arXiv:0805.1726 [gr-qc]].

\bibitem{NojiriOdintsov2011}
S.~Nojiri and S.~D.~Odintsov,
Phys.\ Rept.\ \textbf{505}, 59 (2011).
[arXiv:1011.0544 [gr-qc]].

\bibitem{Capozziello2011}
S.~Capozziello and M.~De Laurentis,
Phys.\ Rept.\ \textbf{509}, 167 (2011).
[arXiv:1108.6266 [gr-qc]].

\bibitem{BeltranJimenez2018}
J.~Beltr\'{a}n Jim\'{e}nez, L.~Heisenberg, and T.~Koivisto,
Phys.\ Rev.\ D \textbf{98}, 044048 (2018).
[arXiv:1710.03116 [gr-qc]].

\bibitem{Li2004}
M.~Li,
Phys.\ Lett.\ B \textbf{603}, 1 (2004).
[arXiv:hep-th/0403127 [hep-th]].

\bibitem{WeiCai2008}
H.~Wei and R.-G.~Cai,
Phys.\ Lett.\ B \textbf{660}, 113 (2008).
[arXiv:0708.0884 [gr-qc]].

\bibitem{Morris1988}
M.~S. Morris and K.~S. Thorne,
Am.\ J.\ Phys.\ \textbf{56}, 395 (1988).

\bibitem{Visser1995}
M.~Visser,
\textit{Lorentzian Wormholes: From Einstein to Hawking}
(AIP Press, Woodbury, NY, 1995).

\bibitem{Hochberg1993}
D.~Hochberg and T.~W. Kephart,
Phys.\ Rev.\ Lett.\ \textbf{70}, 2665 (1993).

\bibitem{Cataldo2008}
M.~Cataldo, S.~Liempi, and P.~Rodr\'{i}guez,
Phys.\ Lett.\ B \textbf{662}, 314 (2008).

\bibitem{Pavlov2014}
M.~Pavlov, A.~Toporensky, and O.~Zaslavskii,
Phys.\ Rev.\ D \textbf{89}, 104054 (2014).

\bibitem{Lobo2009}
F.~S.~N.~Lobo,
Phys.\ Rev.\ D \textbf{75}, 064027 (2007).
[arXiv:gr-qc/0701133 [gr-qc]].

\bibitem{Elizalde2018}
E.~Elizalde and M.~Khurshudyan,
Phys.\ Rev.\ D \textbf{99}, 024051 (2019).
[arXiv:1811.03555 [gr-qc]].

\bibitem{Nojiri2019}
S.~Nojiri, S.~D.~Odintsov, and V.~Faraoni,
Phys.\ Rev.\ D \textbf{98}, 124024 (2018).
[arXiv:1810.02680 [gr-qc]].

\bibitem{ForemanMackey2013}
D.~Foreman-Mackey, D.~W.~Hogg, D.~Lang, and J.~Goodman,
Publ.\ Astron.\ Soc.\ Pac.\ \textbf{125}, 306 (2013).
[arXiv:1202.3665 [astro-ph.IM]].

\bibitem{Lewis2019}
A.~Lewis,
GetDist: a Python package for analysing Monte Carlo samples (2019).
[arXiv:1910.13970 [astro-ph.IM]].

\bibitem{Jimenez2002}
R.~Jim\'{e}nez and A.~Loeb,
Astrophys.\ J.\ \textbf{573}, 37 (2002).
[arXiv:astro-ph/0106145 [astro-ph]].

\bibitem{Moresco2020}
M.~Moresco \textit{et al.},
Astrophys.\ J.\ Suppl.\ \textbf{260}, 1 (2022).
[arXiv:2201.07241 [astro-ph.CO]].

\bibitem{Riess2004}
A.~G. Riess \textit{et al.},
Astrophys.\ J.\ \textbf{607}, 665 (2004).
[arXiv:astro-ph/0402512 [astro-ph]].

\bibitem{Farooq2017}
O.~Farooq, F.~R. Madiyar, S.~Crandall, and B.~Ratra,
Astrophys.\ J.\ \textbf{835}, 26 (2017).
[arXiv:1601.01984 [astro-ph.CO]].

\bibitem{Riess2022}
A.~G. Riess \textit{et al.},
Astrophys.\ J.\ Lett.\ \textbf{934}, L7 (2022).
[arXiv:2112.04510 [astro-ph.CO]].

\bibitem{Nojiri2005}
S.~Nojiri and S.~D.~Odintsov,
Phys.\ Rev.\ D \textbf{71}, 123509 (2005).
[arXiv:hep-th/0504052 [hep-th]].

\bibitem{DeFelice2010}
A.~De Felice and S.~Tsujikawa,
Living Rev.\ Rel.\ \textbf{13}, 3 (2010).
[arXiv:1002.4928 [gr-qc]].

\bibitem{Cai2016}
Y.-F.~Cai, S.~Capozziello, M.~De Laurentis, and E.~N.~Saridakis,
Rept.\ Prog.\ Phys.\ \textbf{79}, 106901 (2016).
[arXiv:1511.07586 [gr-qc]].

\bibitem{Brevik2005}
I.~Brevik, O.~Gorbunova, and Y.~A.~Shaido,
Int.\ J.\ Mod.\ Phys.\ D \textbf{14}, 1899 (2005).
[arXiv:gr-qc/0508038 [gr-qc]].

\bibitem{Cataldo2005}
M.~Cataldo, N.~Cruz, and S.~Lepe,
Phys.\ Lett.\ B \textbf{619}, 5 (2005).
[arXiv:hep-th/0506153 [hep-th]].

\bibitem{Kamenshchik2001}
A.~Kamenshchik, U.~Moschella, and V.~Pasquier,
Phys.\ Lett.\ B \textbf{511}, 265 (2001).
[arXiv:gr-qc/0103004 [gr-qc]].

\bibitem{Bento2002}
M.~C. Bento, O.~Bertolami, and A.~A.~Sen,
Phys.\ Rev.\ D \textbf{66}, 043507 (2002).
[arXiv:gr-qc/0202064 [gr-qc]].

\bibitem{Sola2013}
J.~Sol\`{a},
J.\ Phys.\ Conf.\ Ser.\ \textbf{453}, 012015 (2013).
[arXiv:1306.1527 [gr-qc]].

\bibitem{Verlinde2011}
E.~P. Verlinde,
JHEP \textbf{1104}, 029 (2011).
[arXiv:1001.0785 [hep-th]].

\bibitem{Padmanabhan2010}
T.~Padmanabhan,
Rept.\ Prog.\ Phys.\ \textbf{73}, 046901 (2010).
[arXiv:0911.5004 [gr-qc]].

\bibitem{Bamba2012}
K.~Bamba, S.~Capozziello, S.~Nojiri, and S.~D.~Odintsov,
Astrophys.\ Space Sci.\ \textbf{342}, 155 (2012).
[arXiv:1205.3421 [gr-qc]].

\bibitem{Nojiri2017}
S.~Nojiri, S.~D.~Odintsov, and V.~K.~Oikonomou,
Phys.\ Rept.\ \textbf{692}, 1 (2017).
[arXiv:1705.11098 [gr-qc]].

\bibitem{Frieman2008}
J.~A.~Frieman, M.~S.~Turner, and D.~Huterer,
Ann.\ Rev.\ Astron.\ Astrophys.\ \textbf{46}, 385 (2008).
[arXiv:0803.0982 [astro-ph]].

\bibitem{Weinberg2013}
D.~H.~Weinberg, M.~J.~Mortonson, D.~J.~Eisenstein, C.~Hirata,
A.~G.~Riess, and E.~Rozo,
Phys.\ Rept.\ \textbf{530}, 87 (2013).
[arXiv:1201.2434 [astro-ph.CO]].

\bibitem{Poisson1995}
E.~Poisson and M.~Visser,
Phys.\ Rev.\ D \textbf{52}, 7318 (1995).
[arXiv:gr-qc/9506083 [gr-qc]].

\bibitem{Sushkov2005}
S.~V.~Sushkov,
Phys.\ Rev.\ D \textbf{71}, 043520 (2005).
[arXiv:gr-qc/0502084 [gr-qc]].

\bibitem{Lobo2005}
F.~S.~N.~Lobo,
Phys.\ Rev.\ D \textbf{71}, 084011 (2005).
[arXiv:gr-qc/0502099 [gr-qc]].

\bibitem{Poisson1995b}
P.~R.~Brady, J.~Louko, and E.~Poisson,
Phys.\ Rev.\ D \textbf{44}, 1891 (1991).

\bibitem{Eiroa2004}
E.~F.~Eiroa and G.~E.~Romero,
Gen.\ Rel.\ Grav.\ \textbf{36}, 651 (2004).
[arXiv:gr-qc/0303093 [gr-qc]].

\bibitem{Teo1998}
E.~Teo,
Phys.\ Rev.\ D \textbf{58}, 024014 (1998).
[arXiv:gr-qc/9803098 [gr-qc]].

\bibitem{Bronnikov2013}
K.~A.~Bronnikov and A.~A.~Starobinsky,
Mod.\ Phys.\ Lett.\ A \textbf{24}, 1559 (2009).
[arXiv:0904.0698 [gr-qc]].

\bibitem{Boehmer2012}
C.~G.~B\"{o}hmer, T.~Harko, and F.~S.~N.~Lobo,
Phys.\ Rev.\ D \textbf{85}, 044033 (2012).
[arXiv:1110.5756 [gr-qc]].

\bibitem{Safonova2001}
M.~Safonova, D.~F.~Torres, and G.~E.~Romero,
Phys.\ Rev.\ D \textbf{65}, 023001 (2001).
[arXiv:gr-qc/0106094 [gr-qc]].

\bibitem{Cramer1995}
J.~G.~Cramer, R.~L.~Forward, M.~S.~Morris, M.~Visser, G.~Benford, and
G.~A.~Landis,
Phys.\ Rev.\ D \textbf{51}, 3117 (1995).
[arXiv:astro-ph/9409051 [astro-ph]].

\bibitem{Perivolaropoulos2022}
L.~Perivolaropoulos and F.~Skara,
New Astron.\ Rev.\ \textbf{95}, 101659 (2022).
[arXiv:2105.05208 [astro-ph.CO]].

\bibitem{DiValentino2021}
E.~Di Valentino \textit{et al.},
Class.\ Quant.\ Grav.\ \textbf{38}, 153001 (2021).
[arXiv:2103.01183 [astro-ph.CO]].

\bibitem{Abbott2016}
B.~P.~Abbott \textit{et al.} (LIGO Scientific and Virgo Collaborations),
Phys.\ Rev.\ Lett.\ \textbf{116}, 061102 (2016).
[arXiv:1602.03837 [gr-qc]].

\bibitem{Akiyama2019}
K.~Akiyama \textit{et al.} (Event Horizon Telescope Collaboration),
Astrophys.\ J.\ Lett.\ \textbf{875}, L1 (2019).
[arXiv:1906.11238 [astro-ph.HE]].

\bibitem{Ellis1973}
H.~G.~Ellis,
J.\ Math.\ Phys.\ \textbf{14}, 104 (1973).

\bibitem{Bronnikov1973}
K.~A.~Bronnikov,
Acta Phys.\ Polon.\ B \textbf{4}, 251 (1973).

\bibitem{Bambi2013}
C.~Bambi,
Phys.\ Rev.\ D \textbf{87}, 107501 (2013).
[arXiv:1304.5691 [gr-qc]].

\bibitem{Harko2009}
T.~Harko, Z.~Kov\'{a}cs, and F.~S.~N.~Lobo,
Phys.\ Rev.\ D \textbf{79}, 064001 (2009).
[arXiv:0901.3926 [gr-qc]].

\bibitem{Kar1994}
S.~Kar,
Phys.\ Rev.\ D \textbf{49}, 862 (1994).

\bibitem{Kar1996}
S.~Kar and D.~Sahdev,
Phys.\ Rev.\ D \textbf{53}, 722 (1996).
[arXiv:gr-qc/9506094 [gr-qc]].

\bibitem{Anchordoqui1997}
L.~A.~Anchordoqui, S.~E.~Perez Bergliaffa, and D.~F.~Torres,
Phys.\ Rev.\ D \textbf{55}, 5226 (1997).
[arXiv:gr-qc/9610070 [gr-qc]].

\bibitem{Jamil2010}
M.~Jamil, P.~K.~F.~Kuhfittig, F.~Rahaman, and Sk.~A.~Rakib,
Eur.\ Phys.\ J.\ C \textbf{67}, 513 (2010).
[arXiv:0906.2142 [gr-qc]].

\bibitem{Bhar2016}
P.~Bhar, F.~Rahaman, T.~Manna, and A.~Banerjee,
Eur.\ Phys.\ J.\ C \textbf{76}, 708 (2016).
[arXiv:1611.01340 [gr-qc]].

\bibitem{Bohmer2010CQG}
C.~G.~B\"{o}hmer, T.~Harko, and F.~S.~N.~Lobo,
Class.\ Quant.\ Grav.\ \textbf{27}, 185013 (2010).
[arXiv:1007.1953 [gr-qc]].

\bibitem{Lobo2008CQG}
F.~S.~N.~Lobo,
Class.\ Quant.\ Grav.\ \textbf{25}, 175006 (2008).
[arXiv:0801.4401 [gr-qc]].

\bibitem{Koussour2023}
M.~Koussour, N.~Myrzakulov, Alnadhief H.~A.~Alfedeel, and A.~Abebe,
Prog.\ Theor.\ Exp.\ Phys.\ \textbf{2023}, 113E01 (2023).
[arXiv:2309.10101 [gr-qc]].

\bibitem{Myrzakulov2023}
N.~Myrzakulov, M.~Koussour, Alnadhief H.~A.~Alfedeel, and E.~I.~Hassan,
Prog.\ Theor.\ Exp.\ Phys.\ \textbf{2023}, 093E02 (2023).
[arXiv:2306.13612 [gr-qc]].

\end{thebibliography}
\end{document}